\documentclass[useAMS]{mn2e}
\usepackage{amssymb,amsmath}
\usepackage{graphicx}
\usepackage{lscape}

\title[PG 0052+251]
{More Evidence for the Intermediate Broad Line Region of the Mapped AGN 
PG 0052+251}
\author[Zhang X.-G.]
       {Xue-Guang Zhang$^{1,2}$\\
       $^1$Purple Mountain Observatory, Chinese Academy of Sciences,
             2 Beijing XiLu, NanJing, JiangSu, 210008, P. R. China \\
       $^2$Chinese Center for Antarctic Astronomy, NanJing,
             JiangSu, 210008, P. R. China}

\date{}
\def\LaTeX{L\kern-.36em\raise.3ex\hbox{a}\kern-.15em
    T\kern-.1667em\lower.7ex\hbox{E}\kern-.125emX}
\begin{document}

\pagerange{\pageref{firstpage}--\pageref{lastpage}} \pubyear{}
\maketitle
\label{firstpage}

\begin{abstract}
   In the manuscript, the properties of the proposed intermediate 
BLR are checked for the mapped AGN PG 0052+251. With the considerations 
of the apparent effects of the broad He~{\sc ii} line on the observed 
broad H$\beta$ profile, the line parameters (especially the line width 
and the line flux) of the observed broad H$\alpha$ and the broad H$\beta$ 
are carefully determined. Based on the measured line parameters, the 
model with two broad components applied for each observed broad balmer 
line is preferred, and then confirmed by the calculated much different 
time lags for the inner/intermediate broad components and the 
corresponding virial BH masses ratio determined by the properties of 
the inner and the intermediate broad components. Then, the correlation 
between the broad line width and the broad line flux is checked for the 
two broad components: one clearly strong negative correlation for the 
inner broad component, but one positive correlation for the intermediate 
broad component. The different correlations for the two broad components 
strongly support the intermediate BLR of PG0052+251. 
\end{abstract}

\begin{keywords}
Galaxies:Active -- Galaxies:nuclei -- Galaxies:quasars:Emission lines
-- Galaxies: Individual: PG 0052+251
\end{keywords}

\section{Introduction}
   PG 0052+251 is one well studied mapped blue quasar (Bentz et al. 2009, 
Chelouche \& Daniel 2012, Collin et al. 2006, Kaspi et al. 2000, 2005, 
Peterson et al. 2004, Zu et al. 2011, Zhang 2011b). Based on the 
measured size of the BLR (Broad emission Line Region) and the line width 
of the broad $H\alpha$ and the broad H$\beta$ in the literature, we 
(Zhang 2011b, Paper I) have shown that the blue quasar PG 0052+251 is 
one special object in the plane of 
$(\frac{\sigma(H\beta)}{\sigma(H\alpha)})^2$  versus 
$\frac{R_{BLR, H\alpha}}{R_{BLR, H\beta}}$, where $\sigma$ and $R_{BLR}$ 
mean the measured broad line width based on the mean/rms spectra and the 
size of the BLR determined by the reverberation mapping technique 
(Blandford \& Mckee 1992, Peterson 1993, Peterson \& Horne 2004), 
because of the much different virial black hole masses determined by 
the parameters of the broad H$\beta$ and the broad H$\alpha$. So that, 
in the Paper I, we have reported that it is more appropriate for 
PG 0052+251 to describe the observed broad H$\alpha$ by two broad 
components rather than by one broad component, and then reported the 
strong evidence for the intermediate BLR with the size about 700 
light-days, besides the common BLR with the size about 100-200 
light-days as discussed in Kaspi et al. (2000, 2005), 
Peterson et al. (2004) etc.. The method in Paper I to determine the 
intermediate BLR of PG 0052+251 is much different from the other methods 
by emission line fitted results, such as in Bon et al. (2009), 
Hu et al. (2012), Shapovalova et al. (2012), Zhu et al. (2009) etc. As 
the first reported mapped AGN with the reliable intermediate BLR with 
measured size by the mapping technique, we will further discuss whether 
there are different intrinsic properties for the common inner BLR and 
the intermediate BLR of PG 0052+251.

   In Paper I, we have discussed that the intermediate BLR is not 
the extended part of the common inner BLR, i.e, there is enough physical 
geometrical distance between the common inner BLR and the intermediate 
BLR. Therefore, it will be interesting to check whether the common inner 
BLR and the proposed intermediate BLR have much different kinematic 
properties (such as much different properties of kepler velocities of 
the line clouds in the two regions), which is the main objective of the 
manuscript. In other words, we should check the properties of the line 
parameters (line width tracing properties of kepler velocity, and line 
flux tracing distance between the line region and the central black 
hole) of the broad optical balmer lines from the inner BLR and from the 
intermediate BLR for PG 0052+251.

   It is very difficult to clearly reconstruct the detailed kinematic  
structures of the inner BLR and the intermediate BLR, through the 
sparse and incomplete observational data series of PG 0052+251. 
Therefore, in the manuscript, we only check the simple correlations  
between the line width and the line flux of the broad lines from the two 
line regions. Surely, the correlation should be strongly negative, under 
the Virialization assumption (Bennert et al. 2011, Collin et al. 2006, 
Down et al. 2010, Greene \& Ho 2004, 2005, Kelly \& Bechtold 2007, 
Marziani et al. 2003, Netzer \& Marziani 2010, Onken et al. 2004, 
Park et al. 2012, Peterson et al. 2004, Peterson 2010, Rafiee \& Hall 2011, 
Shen \& Liu 2012, Sluse et al. 2011, Sulentic et al., 2000,
Vestergaard 2002):
\begin{equation}
M_{BH}\propto R_{BLR}\times\sigma^2\propto L^{0.5}\times\sigma^2
\end{equation}
However, besides the commonly expected negative correlation, we 
(Zhang 2013a) have recently reported that for the well-known mapped 
double-peaked emitter (AGN with double-peaked broad low-ionization 
emission lines) 3C390.3 (Dietrich et al. 1998, 2012, Eracleous et al. 1995, 
1997, Flohic \& Eracleous 2008, Popovic et al. 2011, Sergeev et al. 2011, 
Shapovalova et al. 2001, Zhang 2011a, 2013b), the correlation is positive 
for the broad optical balmer lines, which should further indicates the 
accretion disk origination for the observed broad lines. Therefore, 
if the correlations are much different for the broad lines from the inner 
BLR and from the intermediate BLR for PG 0052+251, we should confirm 
that the kinematic properties are intrinsically different for the two line 
regions, and should give some further structure information about the 
intermediate BLR. 

   This manuscript is organized as follows. Section 2 shows the main 
results, including our procedure to measure the line parameters of the 
broad balmer lines, and the line parameters correlations of the broad 
lines. Section 3 gives the discussions and conclusion. Moreover, the 
redshift $z=0.155$ has been accepted for the PG 0052+251\ in the manuscript.

\section{Main Results}

    In Paper I (Zhang 2011b), we have shown that it is much preferred 
to describe the broad observed H$\alpha$ by two broad components. Then, 
PG 0052+251 will have the reasonable location in the plane of 
$(\frac{\sigma(H\beta)}{\sigma(H\alpha)})^2$ versus 
$\frac{R_{BLR, H\alpha}}{R_{BLR, H\beta}}$. However, in Paper I, 
only the mean values and the statistical results are discussed about 
the line parameters of the inner broad H$\alpha$ and the intermediate 
broad H$\alpha$ of PG 0052+251. Here, we will show some more detailed 
results and further discussions about the line parameters of 
both the broad H$\alpha$ and the broad H$\beta$.

    In the manuscript, we consider the observational data and the 
spectra of PG 0052+251 collected from Kaspi et al. (2000) 
(http://wise-obs.tau.ac.il/\~{}shai/PG/). The 53 spectra have both 
apparent broad H$\alpha$ and apparent broad H$\beta$ observed from 
16th Oct. 1991 to 27th Sep. 1998, and  have been well binned into 
1$\AA$ per pixel. Then, the line properties of the broad H$\alpha$ and 
the broad H$\beta$ have been checked for PG 0052+251, with $S/N=10$ 
having been accepted for the collected spectra as discussed in 
Kaspi et al. (2000).

\subsection{Effects of the broad He~{\sc ii}$\lambda4687\AA$ on the 
properties of the Observed Broad H$\beta$ of PG 0052+251}

    As what have been done in Kaspi et al. (2000), Peterson et 
al. (2004) and Zhang (2011b), effects of the optical Fe~{\sc ii} lines 
and He~{\sc ii}$\lambda4686\AA$ line have been totally ignored. 
However, we can find that the apparent He~{\sc ii} lines have strong 
effects on the line profile of the observed broad H$\beta$. Certainly, 
we will find that there are much weak optical Fe~{\sc ii} lines in 
the spectra of PG 0052+251. And moreover, the weak Fe~{\sc ii} lines 
can be well described and removed by our following procedure. Thus, 
besides the effects of the He~{\sc ii} line in the manuscript, there 
are no further discussions for the effects of the Fe~{\sc ii} lines.

\begin{figure*}
\centering\includegraphics[width = 8cm,height=5cm]{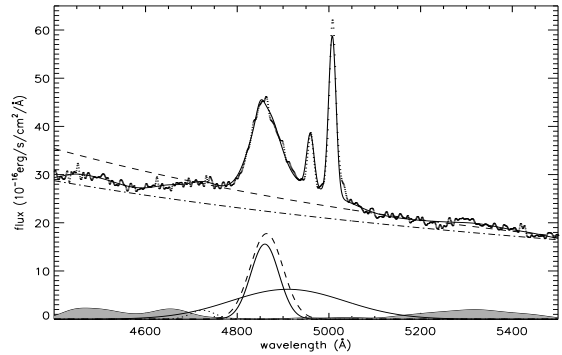}
\caption{Effects of the broad He~{\sc ii} line on the line profile of 
the H$\beta$. The observed spectrum is shown in thin dotted line, 
the best fitted result is shown in the thick solid line, the 
dot-dashed line and the thick dashed line under the spectrum are for the 
power law components with and without the considerations of the broad 
He~{\sc ii} line. The shadow area near the bottom shows the weak 
Fe~{\sc ii} components, the dashed line  near the bottom shows 
the gaussian broad component for the observed broad H$\beta$ without the
consideration of the broad He~{\sc ii}, the two solid lines near the 
bottom show the two gaussian broad components for the observed broad 
H$\beta$ with the consideration of the broad He~{\sc ii}, the 
dotted line near the bottom shows the broad He~{\sc ii} component.
}
\label{heii}
\end{figure*}

  In Kaspi et al. (2000), Peterson et al. (2004) and in Zhang (2011b), 
the AGN continuum emission underneath the observed H$\beta$ is 
determined by the two continuum windows with the rest-wavelength from 
4690\AA to 4750\AA and from 5115\AA to 5175\AA, without the considerations 
of the effects of the weak Fe~{\sc ii} and He~{\sc ii} lines. It is clear 
that the blue window used to determine the continuum is on the shoulder 
of the broad He~{\sc ii} line, which will lead to much steeper 
determined AGN continuum, and lead to much weak broad wings of the 
broad H$\beta$. Figure~\ref{heii} shows the effects of the broad 
He~{\sc ii} line on the the determined AGN continuum, and then the 
effects on the line parameters of the H$\beta$. In the figure, the 
spectrum of PG 0052+251 observed on 14th, Nov. 1993 is shown, because of 
the more apparent He~{\sc ii} and optical Fe~{\sc ii} lines in the 
spectrum. The spectral properties are discussed twice as follows. 
On the one hand, the spectral lines are fitted within the wavelength 
range from 4690\AA to 5175\AA without the consideration of the 
He~{\sc ii} line,  as what have been done in the literature. On the 
other hand, the spectral lines are fitted within the wavelength range 
from 4400\AA to 5500\AA with the consideration of the He~{\sc ii} line 
and the weak optical Fe~{\sc ii} lines. It is clear that there are 
strong effects of the He~{\sc ii} line on the determined AGN continuum 
(the dashed and the dot-dashed lines in the Figure~\ref{heii}).

   Then, the line profile of the broad H$\beta$ is roughly checked. 
More detailed discussions about the fitting procedure for the emission 
lines could be found in the following subsection. If the spectral lines 
are considered within the narrower wavelength range from 4690\AA to 5175\AA, 
the broad H$\beta$ can be well described by one broad gaussian function 
(the dashed line near the bottom in the Figure~\ref{heii}). However, 
the broad H$\beta$ can be well described by two broad gaussian functions
( the solid lines near the bottom in the Figure~\ref{heii}), if the 
spectral lines are considered within the wider wavelength range from 
4400\AA to 5500\AA. Actually, the more recent optical Fe~{\sc ii} template 
discussed in the Kovacevic et al. (2010) has been included in our 
procedure, in order to clearly remove the probable effects of the optical 
Fe~{\sc ii} lines (the shadow areas near the bottom in the 
Figure~\ref{heii}). Based on the results in the Figure~\ref{heii}, we
can find that the Fe~{\sc ii} lines are much weak and have few effects on 
our results about the line parameters of the broad H$\beta$.

    It is clear that there are apparent and strong effects of the 
He~{\sc ii} line on the final results, especially on the AGN continuum 
and on the wings of the broad H$\beta$. Thus, the effects of the 
He~{\sc ii} line should be well considered. And therefore, in the following 
procedure to fit the spectral lines around the H$\beta$, the wider 
wavelength range from 4400\AA to 5500\AA is accepted, rather than the 
narrower wavelength range from 4690\AA to 5175\AA.

\subsection{Line Parameters of the broad H$\beta$ and the broad H$\alpha$}

    In the manuscript, in order to obtain more reliable line 
parameters, the lines around the H$\alpha$ (the broad and the narrow 
H$\alpha$ and the [N~{\sc ii}]$\lambda6548,6583$\AA doublet) and around 
the H$\beta$ (the broad and the narrow H$\beta$, the broad He~{\sc ii} 
line, the [O~{\sc iii}]$\lambda4959,5007$\AA doublet and the optical 
Fe~{\sc ii} line) are fitted simultaneously within the wavelength ranges 
from 4400\AA to 5500\AA for the lines around the H$\beta$ and the 
ranges from 6300\AA to 6900\AA for the lines around the H$\alpha$.
%Here, we should note that there are no considerations about the 
%atmospheric absorption features, because the features have few effects 
%on our results as the discussions in the appendix.

    In the manuscript, two different models are considered to describe 
the observed broad balmer lines: the model with two broad 
gaussian functions applied for each observed broad balmer line, and 
the other model with one broad gaussian function applied for each observed 
broad balmer line. Besides the two models for the broad balmer lines, 
the narrow lines are described by narrow gaussian functions with similar 
line profiles, i.e., they have the same emission line redshifts, the 
same line width. And moreover, the [O~{\sc iii}] ([N~{\sc ii}]) doublet 
has the fixed theoretical intensity ratio. Furthermore, one broad 
gaussian function is applied for the He~{\sc ii} line. Then, two power 
law functions are applied for the continuum under the H$\beta$ and the 
continuum under the H$\alpha$ ($f_{\lambda}\propto\lambda^{\alpha}$).

   Moreover, when two broad gaussian functions are applied for each observed 
broad balmer line, the following restrictions are set: 
\begin{equation}
\begin{split}
\frac{\sigma(H\alpha_1)}{\sigma(H\alpha_2)} &= 
\frac{\sigma(H\beta_1)}{\sigma(H\beta_2)} \\
z(H\beta_1) &= z(H\alpha_1)\\
z(H\beta_2) & = z(H\alpha_2)
\end{split}
\end{equation}
where $\sigma$ and $z$ mean the broad line width (the second moment as 
discussed in Peterson et al. 2004) and the broad emission line redshift  
of the corresponding broad component, the suffixes '1' and '2' represent 
the broad components from the corresponding broad line regions ('1' for 
the inner broad component and '2' for the intermediate broad component). 
%The first formula in the equation indicates the line width ratios are 
%same for the two broad components of the broad H$\alpha$ and the broad H$\beta$. The 
%other two formula in the equation above indicates the broad components of 
%the H$\alpha$ and the H$\beta$ from the corresponding region has the same 
%center wavelengths. 
The restrictions can be reasonably accepted under the following considerations. 

    If the two broad components were physically true for the broad 
H$\alpha$ and the broad H$\beta$, the result could be expected under 
the virialization assumption that
\begin{equation}
\frac{R_{BLR}(H\alpha_1)\times\sigma(H\alpha_1)}{R_{BLR}(H\alpha_2)\times\sigma(H\alpha_2)} = \frac{R_{BLR}(H\beta_1)\times\sigma(H\beta_1)}{R_{BLR}(H\beta_2)\times\sigma(H\beta_2)}
\end{equation} 
where $R_{BLR}$ means the distance between the broad line region and  
the central black hole. Once the results 
$R_{BLR}(H\alpha)\sim R_{BLR}(H\beta)$ are accepted, we can find the 
restriction for the line width ratio in the equation (2). 
And moreover, once we accepted that the inner (intermediate) 
broad components of the H$\alpha$ and the H$\beta$ from the same physical 
emission region, the restriction on the broad line redshift in equation (2) 
can be naturally accepted.

\begin{figure*}
\centering\includegraphics[width = 10cm,height=6cm]{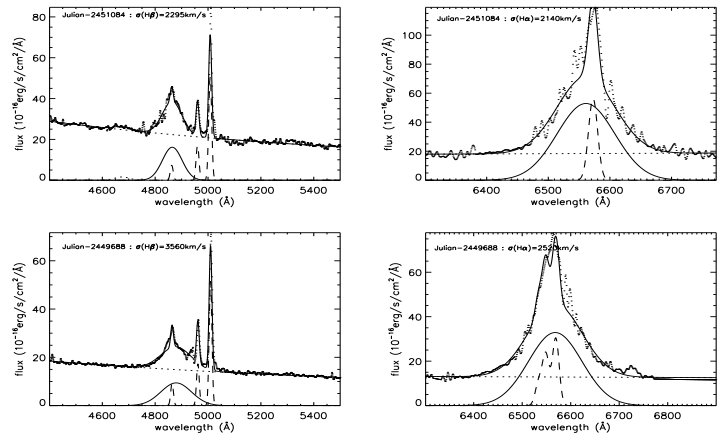}
\caption{Two examples for the best fitted results for the lines 
around the H$\beta$ (left panels) and the H$\alpha$ (right panels), 
under the model with one broad component applied for each observed broad 
balmer line. The thin dotted lines are for the observed spectra, the 
thick solid lines are for the best fitted results. Near the bottom, 
the solid line and the dashed line are for the broad components and the 
other narrow lines. In the left panels, the dotted line around the 4680\AA 
near the bottom shows the probable He~{\sc ii} line. Then, in each panel, 
the MJD and the line width of the broad line are marked.
}
\label{hab1_fit}
\end{figure*}

\begin{figure*}
\centering\includegraphics[width = 12cm,height=5cm]{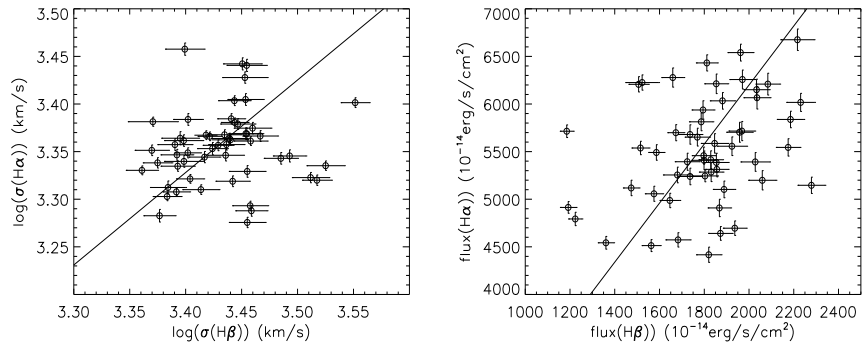}
\caption{The broad line width and broad line flux correlations between 
the broad H$\alpha$ and the broad H$\beta$ under the model with one broad 
component applied for each observed broad balmer line. The line in the 
left panel is $\sigma(H\alpha)=0.97\sigma(H\beta)$, the line in the 
right panel is $flux(H\alpha) = 3.1\times flux(H\beta)$.
}
\label{hab1}
\end{figure*}

   According to the models above and the corresponding restrictions,  
the spectral lines can be well fitted through the Levenberg-Marquardt 
method. Then, the results are firstly checked under the model with one 
broad gaussian function applied for each observed broad balmer line.   
Here, we do not show the best fitted results for all the 53 observed 
spectra, but two simple examples with the maximum and the minimum 
line widths of the broad H$\beta$ in the Figure~\ref{hab1_fit}. 
The basic correlations of the line parameters of the broad H$\alpha$ and 
the broad H$\beta$ are checked and shown in Figure~\ref{hab1}. It is clear 
that there is no clear broad line width correlation: the spearman 
correlation coefficient is 0.14 with $P_{null}=30\%$, and no clear 
broad line flux correlation: the coefficient is 0.29 with $P_{null}\sim3\%$. 
If the broad H$\alpha$ and the broad H$\beta$ are from the unique region, 
strong broad line flux and broad line width correlations should be 
expected. The much weak correlations shown in the Figure~\ref{hab1} 
indicate single broad component applied for each broad balmer line 
is not so reasonable, and further considerations should be checked. 
%Furthermore, we can find that the variations of the line widths and the 
%line fluxes are large enough to ignore the effects of the parameter 
%uncertainties. %Here, when the line flux is compared, the directly 
%measured values are used, not consider the normalization to the 
%[O~{\sc iii}] line, because  

   Then, the model with two broad components applied for each 
observed broad balmer line is considered, with the restrictions in 
the equation (2). And moreover, by the following three steps, the model 
can be checked and further confirmed. The broad line parameters of the 
broad balmer lines are firstly checked under the model. Then, the F-test 
technique is applied to check which model is preferred for the 
broad balmer lines. Finally, the time-lagged correlations are checked.

   The best fitted results for the spectral lines are shown in the 
Figure~\ref{hab2_fit} under the model with two broad components 
applied for each observed broad balmer line. The measured line 
parameters are listed in the Table 1. Then, we check the correlations 
of the line parameters of the broad components of the balmer lines in 
the Figure~\ref{hab2}. The corresponding correlations of the line 
parameters are apparent and strong (coefficient not less than 
0.7 with $P_{null}$ less than $10^{-8}$, the corresponding 
coefficients and $P_{null}$ are marked in each panel of the figure),
\begin{equation}
\begin{split}
\sigma(H\alpha_1) &= 0.95\times\sigma(H\beta_1)\\  
flux(H\alpha_1) &=3.1\times flux(H\beta_1)\\
\sigma(H\alpha_2) &= 0.93\times\sigma(H\beta_2)\\  
flux(H\alpha_2) &=3.1\times flux(H\beta_2)
\end{split}
\end{equation}
It is clear that the broad line width ratios and the broad line flux 
ratios for the inner/intermediate broad components are more reasonable.

   Moreover, the properties about the relative shifted velocities of the 
two broad components are checked and shown as the open circles in 
the Figure~\ref{vel}. It is clear that the time dependent relative shift 
velocities are strongly linear decreasing (the coefficient is 0.61 with 
$P_{null}\sim10^{-6}$), and the best fitted result (the solid line 
in the figure) can be written as,
\begin{equation}
\frac{W_0(1) - W_0(2)}{\rm km/s} \propto (-0.76\pm0.04)\times (Julian-2448837)
\end{equation}
Furthermore, the Figure~\ref{vel} shows the properties of the relative 
shift velocities of the intermediate broad component relative to the 
narrow [O~{\sc iii}]$\lambda5007$\AA (solid circles in the figure). 
We can find that the intermediate broad components have tiny relative 
shift velocities to the narrow [O~{\sc iii}]. The time dependent rather 
than one randomly distributed relative shift velocities between the 
inner broad and the intermediate broad components strongly indicate the 
two components have physical meanings: one line region having no radial 
components and the other component having apparent radial moving 
contributions. Here, we should note the inner BLR including 
different contributions from different components DOES NOT have the 
similar meaning as the multiple BLRs. in the manuscript and in the 
literature, we accept and define that the inner BLR and the 
intermediate BLR mean they are two physical separated line emission 
regions with apparent physical space between the two regions. In other 
words, the observed complicated spectral broad lines do not indicate 
there are two or multiple BLRs, because the different line components 
were perhaps included in the same line region. Meanwhile, the current 
quality of the spectra can not provide further information to 
discuss whether the radial components are from the other line region 
which has apparent physical space from the inner BLR and/or the 
intermediate BLR. Thus, no further discussions about the radial motions 
are shown in the manuscript.

   Then, the F-test technique is applied to determine which model 
is preferred for the observed broad balmer lines. Based on the best 
fitted results by the two models above, the F values firstly are 
calculated as 
\begin{equation}
F = \frac{(SSE_1 - SSE_2)/(DoF_1-DoF_2)}{SSE_2/(DoF_2)}
\end{equation}
where $SSE$ represents the sum of squared residual for one model, 
the suffix '1' is for the model with one broad component applied for 
each observed broad balmer line, and the suffix '2' is for the model 
with two broad components applied for each observed broad balmer line, 
$DoF$ represents the degree of freedom for one model. Through the 
comparison between the calculated value $F$ by the equation (6) and 
the F-value estimated by the F-distribution with the numerator degrees 
of freedom of $DoF_1 - DoF_2$ and the denominator degrees of freedom 
of $DoF_2$, we can conclude our preferred model. Based on the 
selected wavelength range for the lines around the H$\alpha$ and around 
the H$\beta$ and the model parameters, the values of the degrees of 
freedom for the two models are: $DoF_1=1829$, $DoF_2=1826$. It is clear 
that the F-value by the F-distribution with $p=0.05$ is 2.6 (the IDL 
function $f\_cvf(0.05, 3, DoF_{2}-1)$) based on the numerator and 
denominator degrees of freedom. However, the calculated values $F$ 
listed in the Figure~\ref{hab2_fit} by the equation (6) are much 
larger than 2.6, which strongly indicates that it is more preferred 
to describe each observed broad balmer line by two broad components.

   Furthermore, the time-lagged correlations are checked for the 
two broad components. If the inner and the intermediate broad components 
were physically true and from two independent line regions, there should 
be much different time lags between the variabilities of the AGN 
continuum and the two broad components. And then, based on the 
time-lagged results, we can confirm whether the model dependent line 
parameters are physically reliable or only the mathematical model 
results. Before proceeding further, one point we should note. Because 
of the effects of the He~{\sc ii} lines and the narrow lines around 
the H$\alpha$, it should be not so appropriate to directly use the 
scaled light curves of the H$\alpha$ and the H$\beta$ shown in the 
Kaspi et al. (2000), as what we have done in Paper I. Therefore, the 
light curves of the two broad components of the H$\beta$ (the H$\alpha$) 
are determined with the assumption that the [O~{\sc iii}]$\lambda5007$\AA 
flux is constant,
\begin{equation}
flux = flux(obs) \times \frac{710\times10^{-16}{\rm erg/s/cm^2}}{flux([OIII]\lambda5007\AA)}
\end{equation} 
where $710\times10^{-16}{\rm erg/s/cm^2}$ is the mean flux of the 
[O~{\sc iii}]$\lambda5007$\AA, $flux(obs)$ is the line flux  for the 
corresponding broad component directly measured through the observed 
spectrum, $flux$ means the corrected line flux. %The fluxes of 
%the [O~{\sc iii}]$\lambda5007$\AA is shown in the Figure~\ref{o3}. 
It is clear that there is no time-dependent trend for the [O~{\sc iii}], 
and the rms variation about the mean is around 10\%, which indicates 
the flux normalization to the [O~{\sc iii}] flux is necessary. 
Moreover, we can find that the center wavelengths of the 
[O~{\sc iii}]$\lambda5007$\AA have tiny shifts, which is perhaps 
due to the large dispersions and spectral resolutions of the original 
spectra of PG 0052+251 (Kaspi et al. 2000). Meanwhile, the center 
wavelength shifts of the [O~{\sc iii}] have few effects on our final 
results about the broad line parameter correlations. Thus, there are 
no further discussions about the shifts in the manuscript.

   Once the observational light curves of the broad components are 
determined, the commonly accepted ICCF technique (Interpolated 
Cross Correlation Function, Gaskell \& Peterson 1987, Koratkar \& Gaskell 1989, 
Koptelova et al. 2006, Peterson 1993) and the more recent SPEAR 
technique (Stochastic Process Estimation for AGN Reverberation based 
on the damped random walk model for AGN variability, Kelly et al. 2009, 
Kozlowski et al. 2010, MacLeod et al. 2010, Zu et al. 2011) are 
applied to check the time lags between the variabilities of the 
continuum and the inner/intermediate broad components. Here, the light 
curve of the continuum is the one collected from Kaspi et al. (2000). 
Moreover, when the commonly used ICCF method is applied, the direct 
interpolating method is not only applied to the measured observational 
light curves (the open circles in the right panels of Figure~\ref{lag_hb} 
and Figure~\ref{lag_ha}) of the broad components and the continuum 
emission, but also applied to the best descriptions of the light curves 
determined by the damped random walk method (the shadow areas in the 
right panels of Figure~\ref{lag_hb} and Figure~\ref{lag_ha}), in order 
to reduce the effects of the large time gaps of the observational light 
curves. Moreover, the two different method (ICCF and SPEAR) 
should lead to more reliable results, if there were similar ICCF 
and SPEAR results.   

    The time-lagged correlations are shown in the Figure~\ref{lag_hb} for 
the broad components of the H$\beta$ and in the Figure~\ref{lag_ha} for 
the broad components of the H$\alpha$. It is clear that there are apparent 
ans different time lags between the broad components and the AGN continuum 
based on the ICCF and SPEAR results, 
\begin{equation}
\begin{split}
R_{BLR}(H\beta_1,ICCF)&\simeq R_{BLR}(H\alpha_1,ICCF)\\
                 &\sim 270\pm80\hspace{2.5mm} {\rm light-days}\\
R_{BLR}(H\beta_1,SPEAR)&\simeq R_{BLR}(H\alpha_1,SPEAR)\\
                 &\sim 211\pm50\hspace{2.5mm}  {\rm light-days}\\
R_{BLR}(H\beta_2,ICCF)&\simeq R_{BLR}(H\alpha_2,ICCF)\\
                 &\sim 1200\pm300\hspace{2.5mm}  {\rm light-days}\\
R_{BLR}(H\beta_2,SPEAR)&\simeq R_{BLR}(H\alpha_2,SPEAR)\\
                 &\sim 1056\pm160 \hspace{2.5mm} {\rm light-days}\\
\end{split}
\end{equation} 
where the uncertainties with $3\sigma$ confidence levels of the 
time lags are determined by the bootstrap method as what we have done 
in Paper I, and the ICCF results are the mean values based on the 
observational light curves and the light curves determined through the 
random walk method. The results strongly support that the two model 
dependent broad components are both mathematically and physically true, and 
have much different structures, and it is more preferred to describe 
the observed broad balmer line by two broad components for PG 0052+251.

\subsection{On the Correlation between the Line Width and the Line Flux}

   Now, based on the reliable measured line parameters and the 
corresponding uncertainties, the correlations of the broad line parameters 
of the two broad components of the balmer lines can be checked. The linear 
correlation coefficients are about -0.63 with 
$P_{null}\sim9\times10^{-7}$, 0.72 with $P_{null}\sim5\times10^{-9}$, 
-0.48 with $P_{null}\sim4\times10^{-4}$ and 0.69 with 
$P_{null}\sim9\times10^{-8}$ for the correlations between the line width 
and the line flux for the inner broad H$\alpha$, for the intermediate 
broad H$\alpha$, for the inner broad H$\beta$ and for the intermediate 
broad H$\beta$ respectively. The strong correlations are shown in 
Figure~\ref{flux_width}. It is clear that there are much different 
properties for the inner and the intermediate broad components of 
PG 0052+251: one clear positive correlation for the intermediate 
broad component, but one clear negative correlation for the inner broad 
component. The best fitted results for the correlations with the 
considerations about the uncertainties in both the coordinates can be 
written as, 
\begin{equation}
\begin{split}
\log(\frac{flux(H\alpha_1)}{10^{-16}{\rm ergs/s/cm^2}}) &=  
   (7.89\pm0.21) \\
   &- (1.27\pm0.06)\times\log(\frac{\sigma(H\alpha_1)}{\rm km/s}) \\
\log(\frac{flux(H\alpha_2)}{10^{-16}{\rm ergs/s/cm^2}}) &=  
   (-2.16\pm0.33) \\
   &+ (1.81\pm0.11)\times\log(\frac{\sigma(H\alpha_2)}{\rm km/s})\\
\log(\frac{flux(H\beta_1)}{10^{-16}{\rm ergs/s/cm^2}}) &=  
   (8.69\pm0.39) \\
   &- (1.65\pm0.12)\times\log(\frac{\sigma(H\beta_1)}{\rm km/s}) \\
\log(\frac{flux(H\beta_2)}{10^{-16}{\rm ergs/s/cm^2}}) &=  
   (-2.85\pm0.41) \\
   &+ (1.85\pm0.13)\times\log(\frac{\sigma(H\beta_2)}{\rm km/s})
\end{split}
\end{equation}
Meanwhile, the corresponding 99.95\% confidence bands for the best 
fitted results above are also shown in Figure ~\ref{flux_width}. 

   Before the end of the subsection, one point we should mote. Based on 
the fitted results shown in the Figure~\ref{hab2_fit} and the listed 
line parameters in the Table 1, we can find that there are five spectra 
of which the narrow H$\alpha$ (and/or [N~{\sc ii}]$\lambda6583\AA$) are 
much stronger, while the narrow H$\beta$ are much weaker: JD-2448837, 
JD-2449219, JD-2449249, JD-2449279 and JD-2450459. We do not know the 
clear reason why the narrow balmer lines are much different from the 
lines in the other spectra. However, we can find that the results have 
few effects on our final results about the broad line parameters 
correlations, because no different re-calculated broad line parameters 
correlations can be found without the considerations of the five spectra 
with the different narrow lines. Without the considerations of the five 
cases, the corresponding correlation coefficients are about -0.62 with 
$P_{null}\sim6\times10^{-6}$, 0.74 with $P_{null}\sim3\times10^{-9}$, 
-0.49 with $P_{null}\sim7\times10^{-3}$ and 0.72 with 
$P_{null}\sim2\times10^{-8}$ for the correlations between the broad line 
width and the broad line flux for the inner broad H$\alpha$,
for the intermediate broad H$\alpha$, for the inner broad H$\beta$ and
for the intermediate broad H$\beta$ respectively. Therefore, no 
apparent effects of the narrow lines in the several special cases 
can be found for our final results. Thus, there are no further 
discussions about the cases with special narrow lines in the manuscript.

   As we commonly know that the properties of the broad line flux can 
be used to trace the distance between the broad line clouds and the 
central black hole, and the properties of the line width can be used to 
trace the kepler velocities of the broad line clouds. The different 
correlations between the line width and the line flux for the inner and 
the intermediate broad components strongly indicate there are much 
different properties of the two broad line regions of PG 0052+251.

\section{Discussions and Conclusions}

    It is clear that based on the widely accepted Virialization 
assumption for the broad line AGN, the negative correlation between the 
line width and the line flux can be expected for individual object. 
Such negative correlation has been found and confirmed for several 
mapped objects, such as NGC5548 (Bentz et al. 2006, Denney et al. 2010,  
Peterson et al. 2004, Zhang 2011a) and some other mapped objects 
discussed in Peterson et al. (2004). Then based on the empirical 
relation to estimate the size of the BLR $R_{BLR}\propto L^{\alpha}$ 
(Bentz et al. 2006, 2009, Denney et al. 2010, Kaspi et al. 2005, 
Greene et al. 2010, Wang \& Zhang 2003), one strongly negative correlation 
between the broad line flux and the broad line width can be expected 
for individual object. Meanwhile, the common viewpoint about gravity 
dominated BLR can be used to naturally explain the negative correlation  
(Kollatschny \& Zetzl1 2011, Krause et al. 2011, Netzer \& Marziani 2010, 
Peterson \& Wandel 1999, and references therein).

   Moreover, we should note that if the common value $\alpha\sim0.5$ 
was accepted for the empirical relation to estimate the BLR size of 
AGN, we should expect the slope of the correlation for the inner broad 
component is about $-4$, which is smaller than the reported slope, 
$\sim-1.5$ shown in Figure~\ref{flux_width} for PG 0052+251. The difference 
can be well and naturally explained by the following reason: the shape 
of the continuum changes as AGNs vary, i.e., variability amplitudes are 
different for UV and optical bands. And moreover, UV flux rather than 
optical flux is a better measure for ionizing flux which is one much 
better indicator for the size of BLR. So that, the different variability 
amplitudes in UV and optical bands lead to some different slope from 
$-4$ for the correlations of the inner broad components of PG 0052+251, 
and smaller variability amplitudes in optical bands lead to the slope 
larger than $-4$.  

   Besides the negative correlation for the inner broad H$\alpha$, 
there is one clear positive correlation between the line width and 
the line flux for the intermediate broad components of PG 0052+251, 
which is against the virialization assumption. However, we (Zhang 2013a) 
have recently reported that there is one positive correlation 
between the line width and the line flux of the broad double-peaked 
H$\alpha$ of the well-known mapped object 3C390.3. Furthermore, we have 
discussed that the positive correlation of broad line parameters could 
be used as one indicator for the accretion disk origination of the 
broad lines. So that, if the accretion disk origination was accepted for 
the intermediate broad components of PG 0052+251, the positive correlation 
could be naturally explained, as we have explained the positive 
correlation for the double-peaked emitter of 3C390.3.  

  Although there are different dependent modes of the size of BLR on the 
line flux for the two components, the correlation between the 
measured size of the BLR (not the line flux) and the line width 
can be checked under the virialization assumption: similar black hole 
masses determined by the properties of the inner BLR and the intermediate 
BLR
\begin{equation}
\begin{split}
\frac{R_{BLR}(H\alpha_1)\times \sigma^2(H\alpha_1)}
 {R_{BLR}(H\alpha_2)\times \sigma^2(H\alpha_2)} \sim 
  \frac{\frac{226}{{\rm light-days}}\times(\frac{3100}{{\rm km/s}})^2}  
  {\frac{1090}{{\rm light-days}}\times(\frac{1400}{{\rm km/s}})^2} =1.01  \\
\end{split}
\end{equation} 
where $\sigma(H\alpha_1)\sim3100{\rm km/s}$ and 
$\sigma(H\alpha_2)\sim1400{\rm km/s}$ are the mean line widths of the 
inner broad and the intermediate broad components of the H$\alpha$, 
and $R_{BLR}$ is the weighted mean value from the ICCF and SPEAR results 
(Equation (8)). Similar result can be confirmed for the two broad 
components of the H$\beta$. The results indicate the measured time lags 
for the inner/intermediate broad components are reliable, and the two 
model-calculated components are not fake. 

  Before the end of the manuscript, there are two points we should 
note. On the one hand, the main objective of the manuscript is to 
provide further evidence for the intermediate BLR of PG 0052+251. 
The much different and reliable correlations for the two components in the 
Figure~\ref{flux_width} provide enough evidence to support our objective. 
Therefore, no further discussions are shown for the detailed structure 
information on the two broad components. The apparent time-dependent 
shifted velocities of the inner broad component (Figure~\ref{vel}) 
indicates probable radial contributions for the component. The positive 
correlation provides the possibility for the accretion disk origination 
of the intermediate broad components.  So far only two individual 
AGNs, 3C390.3\ in our previous paper and PG 0052+251\ in the manuscript, 
have shown the positive correlation between the line width and the line 
flux. Meanwhile, the accretion disk origination for the double-peaked 
broad lines have been widely accepted for the 3C390.3. Therefore, we 
naturally presume but not confirm that the intermediate broad balmer 
lines of PG 0052 have the similar characters as the lines of 3C390.3. 
Surely, more efforts should be done to give the final confirmed conclusions 
about the structures of the two components, which is beyond the scope 
of the manuscript. On the other hand, more reasonable continuum 
windows are selected to determine the AGN continuum especially under 
the H$\beta$ in the manuscript, which leads to some different BLR size 
based on the H$\beta$ variabilities. If the contributions of the narrow 
lines are subtracted, the measured size of the BLR is around $R\sim200$ 
light-days based on the H$\beta$ variabilities, similar with the values 
determined by the H$\alpha$ variabilities. Therefore, the position of 
PG 0052+251 is reasonable in the plane of 
$(\frac{\sigma(H\beta)}{\sigma(H\alpha)})^2$  versus
$\frac{R_{BLR, H\alpha}}{R_{BLR, H\beta}}$. The selected narrower 
wavelength range to determine the continuum lead the inner broad H$\beta$ 
being much weakened. Yet, in other ways, it is effective to select the 
objects with probable intermediate BLR through the properties of 
objects in the plane of $(\frac{\sigma(H\beta)}{\sigma(H\alpha)})^2$  
versus $\frac{R_{BLR, H\alpha}}{R_{BLR, H\beta}}$.

\vspace{8mm} 

   Our final main conclusions are as follows. 
\begin{itemize}
\iffalse
\item Due to the apparent He~{\sc ii} line, the much wider wavelength 
ranges should be selected to determine the AGN continuum under the 
H$\beta$, otherwise, the wide wings of the H$\beta$ should be artificially 
removed.
\fi 
\item The observed broad balmer lines are being fitted by two models. If 
the model with only one simple broad component was applied for each 
observed broad balmer line, the corresponding broad line parameters 
correlations between the broad H$\alpha$ and the broad H$\beta$ are 
much weak (Figure~\ref{hab1}). However, Under model with two broad 
components applied for each observed broad balmer line, the broad 
line width and the broad line flux correlations are much strong and more 
reasonable for the corresponding broad components of the H$\alpha$ 
and the H$\beta$ (Figure~\ref{hab2}). Moreover, the F-test technique 
has been applied to check the two models, and indicates two broad 
components for the broad balmer line (Figure~\ref{hab2_fit}) are preferred. 
Then, the time-lagged correlations have been checked for the two 
components (Figure~\ref{lag_hb} and Figure~\ref{lag_ha}). The much 
different time lags between the two broad components and the 
continuum ensure that the two components are not mathematical model 
dependent components but have physical meanings for different geometric 
structures.
\item Based on the measured line parameters, one positive correlation 
between the line width and the line flux can be found for the 
intermediate broad component, but one negative correlation can be found 
for the inner broad component,  of the balmer lines of PG 0052+251. 
The different correlations strongly support the intermediate BLR of 
PG 0052+251, and clearly indicate the inner BLR and the intermediate 
BLR have much different dynamic/geometric structures. 
\end{itemize}

\section*{Acknowledgements}
ZXG very gratefully acknowledge the anonymous referee for 
giving us constructive comments and suggestions to greatly 
improve our paper. ZXG gratefully acknowledges the support from  
NSFC-11003043 and NSFC-11178003, and gratefully thanks Dr. Kaspi S. to 
provide public observed spectra of PG 0052+251 
(http://wise-obs.tau.ac.il/\~{}shai/PG/).

\clearpage
\begin{landscape}
\begin{table}
%\begin{minipage}{220mm}
\caption{Line Parameters of PG 0052+251}
\begin{tabular}{ccccccccccccc} 
\hline
\multicolumn{1}{c}{JD} &\multicolumn{3}{c}{Inner Broad} & \multicolumn{3}{c}{Intermediate Broad}& \multicolumn{3}{c}{[O~{\sc iii}]} & \multicolumn{3}{c}{' '}\\
  & $W_0$ & $\sigma$ & flux & $W_0$ & $\sigma$ & flux & $W_0$ & $\sigma$ & 
   flux & flux(H$\beta_N$) & flux(H$\alpha_N$) & flux([N~{\sc ii}]) \\
\hline
48837  &  4858.9$\pm$5.0  &  3967$\pm$469  &  8.81$\pm$4.88  &  
          4868.6$\pm$1.7  &  2568$\pm$234  &  14.68$\pm$4.74  &  
          5012.0$\pm$0.3  &  660$\pm$17  &  9.53$\pm$0.28  &  
          0.13$\pm$0.40  &  10.05$\pm$1.17  &     \\
   &  6554.4$\pm$6.6  &  2965$\pm$349  &  20.29$\pm$11.21  &  
      6567.5$\pm$2.3  &  1920$\pm$234  &  33.86$\pm$6.41  &     &     &     &     &     &     \\
48885  &  4900.7$\pm$18.2  &  4249$\pm$640  &  5.46$\pm$2.41  &  
          4853.4$\pm$1.7  &  2591$\pm$177  &  13.95$\pm$2.06  &  
          5007.6$\pm$0.2  &  389$\pm$13  &  6.74$\pm$0.24  &  
          1.13$\pm$0.25  &  5.07$\pm$0.58  &     \\
   &  6619.1$\pm$24.1  &  2828$\pm$640  &  13.98$\pm$3.25  &  
      6555.3$\pm$1.3  &  1724$\pm$95.7  &  35.75$\pm$5.73  &     &     &     &     &     &     \\
48989  &  4893.5$\pm$5.5  &  5253$\pm$365  &  5.51$\pm$0.74  &  
          4864.7$\pm$1.4  &  2406$\pm$107  &  15.45$\pm$0.78  &  
          5008.0$\pm$0.2  &  383$\pm$14  &  6.81$\pm$0.24  &  
          0.67$\pm$0.27  &  3.22$\pm$0.62  &     \\
   &  6600.8$\pm$7.2  &  3972$\pm$227  &  17.48$\pm$2.28  &  
      6561.9$\pm$1.9  &  1819$\pm$107  &  49.05$\pm$1.06  &     &     &     &     &     &     \\
49004  &  4899.3$\pm$3.2  &  4925$\pm$338  &  9.71$\pm$0.88  &  
          4869.0$\pm$1.7  &  2117$\pm$142  &  12.03$\pm$0.77  &  
          5010.7$\pm$0.4  &  611$\pm$22  &  7.45$\pm$0.29  &  
          0.69$\pm$0.46  &  4.14$\pm$0.95  &  7.73$\pm$1.04  \\
   &  6587.8$\pm$3.6  &  3852$\pm$115  &  28.82$\pm$1.87  &  
      6547.1$\pm$2.3  &  1656$\pm$142  &  35.74$\pm$1.04  &     &     &     &     &     &     \\
49011  &  4895.3$\pm$11.1  &  3187$\pm$328  &  3.85$\pm$1.31  &  
          4861.8$\pm$1.9  &  1955$\pm$153  &  11.34$\pm$1.49  &  
          5008.2$\pm$0.2  &  424$\pm$16  &  6.53$\pm$0.24  &  
          0.49$\pm$0.32  &  4.14$\pm$0.61  &     \\
   &  6603.8$\pm$16.1  &  3363$\pm$328  &  16.35$\pm$1.77  &  
      6558.6$\pm$1.4  &  2062$\pm$90  &  48.13$\pm$5.21  &     &     &     &     &     &     \\
49219  &  4875.3$\pm$1.5  &  4012$\pm$219  &  16.38$\pm$0.76  &  
          4863.2$\pm$1.4  &  1398$\pm$136  &  6.286$\pm$0.61  &  
          5006.9$\pm$0.2  &  407$\pm$12  &  6.52$\pm$0.24  &  
          1.18$\pm$0.34  &  14.31$\pm$0.93  &     \\
   &  6550.1$\pm$1.2  &  2297$\pm$48  &  34.81$\pm$1.19  &  
      6533.8$\pm$1.8  &  801$\pm$136  &  13.35$\pm$0.81  &     &     &     &     &     &     \\
49220  &  4892.2$\pm$17.1  &  5728$\pm$1002  &  1.53$\pm$0.58  &  
          4861.8$\pm$1.3  &  2570$\pm$110  &  17.01$\pm$0.71  &  
          5007.6$\pm$0.3  &  629$\pm$18  &  8.18$\pm$0.24  &  
          1.72$\pm$0.38  &     &     \\
   &  6608.3$\pm$23.2  &  4027$\pm$1002  &  4.63$\pm$0.79  &  
      6567.3$\pm$0.7  &  1806$\pm$38  &  51.64$\pm$1.81  &     &     &     &     &     &     \\
49249  &  4893.0$\pm$2.3  &  7634$\pm$388  &  15.89$\pm$0.82  &  
          4865.5$\pm$1.8  &  2252$\pm$159  &  8.91$\pm$0.54  &  
          5008.3$\pm$0.3  &  569$\pm$19  &  6.86$\pm$0.26  &  
          1.54$\pm$0.34  &  13.45$\pm$1.07  &  6.71$\pm$0.62  \\
   &  6565.7$\pm$1.7  &  3341$\pm$80  &  32.45$\pm$1.14  &  
      6528.9$\pm$2.4  &  985$\pm$159  &  18.18$\pm$0.72  &     &     &     &     &     &     \\
49250  &  4871.8$\pm$2.6  &  3916$\pm$338  &  7.39$\pm$1.43  &  
          4865.6$\pm$1.2  &  2022$\pm$130  &  11.88$\pm$1.38  &  
          5006.1$\pm$0.3  &  594$\pm$18  &  8.15$\pm$0.25  &  
          1.13$\pm$0.41  &  1.01$\pm$0.91  &     \\
   &  6567.5$\pm$3.1  &  3223$\pm$237  &  20.71$\pm$3.98  &  
      6559.1$\pm$1.7  &  1664$\pm$130  &  33.28$\pm$1.86  &     &     &     &     &     &     \\
49279  &  4874.1$\pm$2.2  &  4327$\pm$286  &  10.79$\pm$0.95  &  
          4861.2$\pm$1.1  &  1930$\pm$109  &  13.51$\pm$0.92  &  
          5007.7$\pm$0.3  &  584$\pm$19  &  6.73$\pm$0.24  &  
          0.43$\pm$0.47  &  12.22$\pm$1.05  &     \\
   &  6560.6$\pm$2.4  &  3512$\pm$139  &  28.55$\pm$2.25  &  
      6543.3$\pm$1.5  &  1566$\pm$109  &  35.73$\pm$1.25  &     &     &     &     &     &     \\
49282  &  4881.5$\pm$2.3  &  5581$\pm$288  &  6.41$\pm$0.41  &  
          4865.9$\pm$1.2  &  2198$\pm$92  &  11.85$\pm$0.46  &  
          5010.4$\pm$0.2  &  409$\pm$13  &  5.96$\pm$0.19  &  
          0.63$\pm$0.23  &  0.62$\pm$0.53  &     \\
   &  6582.7$\pm$2.6  &  4183$\pm$143  &  22.01$\pm$1.29  &  
      6561.6$\pm$1.6  &  1648$\pm$91  &  40.74$\pm$0.62  &     &     &     &     &     &     \\
49303  &  4884.6$\pm$2.9  &  4903$\pm$316  &  8.85$\pm$0.77  &  
          4862.6$\pm$1.5  &  2292$\pm$132  &  11.25$\pm$0.71  &  
          5008.6$\pm$0.3  &  615$\pm$20  &  7.49$\pm$0.26  &  
          0.85$\pm$0.37  &  4.11$\pm$0.91  &     \\
   &  6585.9$\pm$3.3  &  3485$\pm$113  &  25.24$\pm$2.01  &  
      6556.1$\pm$2.1  &  1629$\pm$132  &  32.10$\pm$0.96  &     &     &     &     &     &     \\
49306  &  4875.7$\pm$3.3  &  4845$\pm$370  &  6.87$\pm$0.96  &  
          4863.1$\pm$1.3  &  2321$\pm$126  &  13.93$\pm$0.95  &  
          5007.6$\pm$0.3  &  595$\pm$18 &  8.37$\pm$0.26  &  
          0.85$\pm$0.39  &  0.63$\pm$0.96  &  0.92$\pm$0.91  \\
   &  6577.9$\pm$3.8  &  3611$\pm$208  &  21.88$\pm$2.99  &  
      6561.0$\pm$1.8  &  1730$\pm$126  &  44.34$\pm$1.28  &     &     &     &     &     &     \\
49325  &  4876.6$\pm$2.2  &  6137$\pm$328  &  18.36$\pm$1.25  &  
          4859.4$\pm$1.9  &  2324$\pm$202  &  8.91$\pm$0.92  &  
          5007.7$\pm$0.3  &  575$\pm$22  &  6.94$\pm$0.31  &  
          1.75$\pm$0.37  &  8.33$\pm$1.21  &  4.34$\pm$0.65  \\
   &  6553.6$\pm$1.4  &  2726$\pm$75  &  36.73$\pm$1.91  &  
      6530.5$\pm$2.6  &  1032$\pm$202  &  17.81$\pm$1.24  &     &     &     &     &     &     \\
49366  &  4876.6$\pm$2.6  &  5219$\pm$397  &  8.08$\pm$1.38  &  
          4875.6$\pm$1.5  &  2590$\pm$135  &  13.70$\pm$1.36  &  
          5007.3$\pm$0.2  &  499$\pm$16  &  7.03$\pm$0.25  &  
          1.16$\pm$0.28  &  5.67$\pm$0.89  &  3.69$\pm$0.74  \\
   &  6549.6$\pm$2.4  &  3207$\pm$218  &  24.77$\pm$4.02  &  
      6548.2$\pm$2.0  &  1591$\pm$135  &  42.04$\pm$1.84  &     &     &     &     &     &     \\
49386  &  4883.7$\pm$4.8  &  4328$\pm$330  &  5.42$\pm$1.02  &  
          4864.6$\pm$1.4  &  2269$\pm$115  &  13.17$\pm$1.01  &  
          5007.3$\pm$0.2  &  393$\pm$12  &  7.01$\pm$0.23  &  
          0.85$\pm$0.26  &  5.26$\pm$0.68  &  2.21$\pm$0.61  \\
   &  6590.4$\pm$6.1  &  3533$\pm$219  &  18.31$\pm$3.43  &  
      6564.6$\pm$1.9  &  1852$\pm$115  &  44.54$\pm$1.36  &     &     &     &     &     &     \\
49542  &  4864.7$\pm$1.4  &  2175$\pm$78  &  14.58$\pm$0.54  &  
          4789.8$\pm$8.5  &  1771$\pm$368  &  1.11$\pm$0.36  &  
          5005.6$\pm$0.1  &  370$\pm$11  &  7.42$\pm$0.22  &  
          1.01$\pm$0.22  &  5.53$\pm$0.48  &     \\
   &  6557.0$\pm$1.5  &  1980$\pm$51  &  43.17$\pm$1.24  &  
      6456.1$\pm$11.1  &  1612$\pm$368  &  3.27$\pm$0.49  &     &     &     &     &     &     \\
49569  &  4869.0$\pm$1.8  &  4559$\pm$251  &  7.59$\pm$0.65  &  
          4864.7$\pm$1.2  &  2040$\pm$97  &  9.47$\pm$0.65  &  
          5007.0$\pm$0.1  &  347$\pm$9  &  7.08$\pm$0.22  &  
          0.74$\pm$0.22  &  2.44$\pm$0.65  &  0.12$\pm$0.54  \\
   &  6567.6$\pm$1.7  &  3369$\pm$141  &  27.27$\pm$2.32  &  
      6561.8$\pm$1.6  &  1508$\pm$97  &  34.04$\pm$0.87  &     &     &     &     &     &     \\
49571  &  4869.1$\pm$1.5  &  3193$\pm$160  &  10.55$\pm$0.97  &  
          4861.5$\pm$1.3  &  1656$\pm$136  &  5.39$\pm$0.94  &  
          5007.7$\pm$0.1  &  346$\pm$9  &  7.12$\pm$0.21  &  
          0.88$\pm$0.21  &  1.91$\pm$0.57  &     \\
   &  6566.0$\pm$1.3  &  2476$\pm$81  &  38.67$\pm$3.44  &  
      6555.8$\pm$1.8  &  1284$\pm$136  &  19.78$\pm$1.27  &     &     &     &     &     &     \\
\hline
\end{tabular}\\
%\begin{minipage}{200mm}
Notice: The first column gives the observational dates (JD-2400000),
the second to forth columns show the line parameters of the inner
broad component: center wavelength $W_0$ in unit of $\AA$, line
width $\sigma$ in unit of ${\rm km/s}$ and line flux in unit of
$10^{-14}{\rm erg/s/cm^2}$, the fifth to seventh columns show the line
parameters of the intermediate broad component, the eighth to tenth 
columns show the line parameters of the narrow [O~{\sc iii}], 
the last three columns show the line fluxes (in unit of 
$10^{-14}{\rm erg/s/cm^2}$) of the narrow H$\beta$, narrow 
H$\alpha$ and [N~{\sc ii}]$\lambda6583\AA$. Because the narrow lines 
have the same line redshift and the same line width (${\rm km/s}$), 
thus we do not list the center wavelengths and the line widths of the 
narrow H$\beta$, the narrow H$\alpha$ and the [N~{\sc ii}] line in the 
table.  The first line of every two rows lists the parameters for the 
two components of the H$\beta$ (second to seventh columns), and the 
second line lists the parameters for the H$\alpha$ (second to seventh 
columns).
%\end{minipage}
\end{table}
\end{landscape}

\clearpage

\begin{landscape}
\begin{table}
\addtocounter{table}{-1}
\caption{continued}
\begin{tabular}{ccccccccccccc}
\hline
\multicolumn{1}{c}{JD} &\multicolumn{3}{c}{Inner Broad} & \multicolumn{3}{c}{Intermediate Broad}& \multicolumn{3}{c}{[O~{\sc iii}]} & \multicolumn{3}{c}{' '}\\
  & $W_0$ & $\sigma$ & flux & $W_0$ & $\sigma$ & flux & $W_0$ & $\sigma$ &
   flux & flux(H$\beta_N$) & flux(H$\alpha_N$) & flux([N~{\sc ii}]) \\
\hline
49596  &  4869.7$\pm$1.7  &  3929$\pm$221  &  7.02$\pm$0.76  &
          4863.8$\pm$1.2  &  1903$\pm$95  &  8.78$\pm$0.75  &
          5008.1$\pm$0.1  &  353$\pm$8  &  7.29$\pm$0.19  &
          0.65$\pm$0.19  &  2.69$\pm$0.55  &     \\
   &  6566.5$\pm$1.7  &  3093$\pm$141  &  25.95$\pm$2.74  &
      6558.5$\pm$1.4  &  1498$\pm$95  &  32.46$\pm$1.01  &     &     &     &     &     &     \\
49597  &  4877.4$\pm$1.9  &  4464$\pm$209  &  6.29$\pm$0.54  &
          4865.5$\pm$1.3  &  2097$\pm$78  &  10.77$\pm$0.55  &
          5009.4$\pm$0.1  &  361$\pm$8  &  6.83$\pm$0.18  &
          0.81$\pm$0.19  &  1.81$\pm$0.54  &     \\
   &  6577.9$\pm$2.1  &  3393$\pm$124  &  24.25$\pm$2.04  &
      6561.8$\pm$1.3  &  1594$\pm$77  &  41.51$\pm$0.75  &     &     &     &     &     &     \\
49598  &  4878.5$\pm$2.2  &  4476$\pm$209  &  6.43$\pm$0.54  &
          4861.8$\pm$1.0  &  2105$\pm$75  &  11.62$\pm$0.55  &
        5006.7$\pm$0.1  &  356$\pm$8  &  6.94$\pm$0.18  &
        0.75$\pm$0.19  &  2.33$\pm$0.59  &  1.01$\pm$0.52  \\
   &  6579.9$\pm$2.6  &  3450$\pm$119  &  23.97$\pm$1.94  &
        6557.4$\pm$1.3  &  1622$\pm$75  &  43.37$\pm$0.75  &     &     &     &     &     &     \\
49661  &  4868.5$\pm$1.3  &  3446$\pm$173  &  6.69$\pm$0.66  &  
	4870.1$\pm$1.2  &  1736$\pm$101  &  5.91$\pm$0.64  &  
	5013.0$\pm$0.1  &  351$\pm$8  &  6.93$\pm$0.17  &  
	0.67$\pm$0.16  &  0.39$\pm$0.57  &     \\
   &  6565.0$\pm$1.2  &  2703$\pm$100  &  32.13$\pm$3.11  &  
	6567.1$\pm$1.4  &  1362$\pm$101  &  28.40$\pm$0.87  &     &     &     &     &     &     \\
49681  &  4877.5$\pm$1.9  &  4561$\pm$215  &  5.38$\pm$0.41  &  
	4864.1$\pm$1.1  &  2102$\pm$87  &  7.85$\pm$0.41  &  
	5010.1$\pm$0.1  &  364$\pm$8  &  6.67$\pm$0.17  &  
	0.78$\pm$0.16  &  2.13$\pm$0.51  &     \\
   &  6580.0$\pm$2.0  &  3357$\pm$108  &  23.83$\pm$1.71  &  
	6562.0$\pm$1.5  &  1547$\pm$87  &  34.75$\pm$0.56  &     &     &     &     &     &     \\
49688  &  4883.9$\pm$2.0  &  5483$\pm$255  &  7.57$\pm$0.51  &  
	4873.8$\pm$1.6  &  2417$\pm$115  &  7.05$\pm$0.47  &  
	5009.5$\pm$0.1  &  318$\pm$7 &  6.86$\pm$0.19  &  
	0.98$\pm$0.14  &  2.04$\pm$0.46  &     \\
   &  6574.1$\pm$1.5  &  3204$\pm$95  &  28.98$\pm$1.73  &  
	6560.5$\pm$2.2  &  1412$\pm$115  &  26.97$\pm$0.63  &     &     &     &     &     &     \\
49719  &  4875.3$\pm$1.8  &  3949$\pm$212  &  8.19$\pm$0.52  &  
	4864.9$\pm$1.5  &  1771$\pm$119  &  4.73$\pm$0.48  &  
	5011.3$\pm$0.1  &  321$\pm$8  &  6.87$\pm$0.19  &  
	0.78$\pm$0.16  &  2.08$\pm$0.46  &     \\
   &  6571.7$\pm$1.3  &  2855$\pm$77 &  33.72$\pm$1.89  &  
	6557.6$\pm$2.1  &  1280$\pm$119  &  19.48$\pm$0.64  &     &     &     &     &     &     \\
49921  &  4870.8$\pm$1.9  &  4081$\pm$263  &  10.03$\pm$1.08  &  
	4863.6$\pm$1.4  &  1978$\pm$133  &  8.719$\pm$1.05  &  
	5008.4$\pm$0.1  &  314$\pm$9  &  6.96$\pm$0.23  &  
	0.77$\pm$0.22  &  0.94$\pm$0.51  &     \\
   &  6568.8$\pm$1.8  &  2903$\pm$130  &  31.05$\pm$3.20  &  
	6559.1$\pm$1.9  &  1408$\pm$133  &  26.99$\pm$1.42  &     &     &     &     &     &     \\
49962  &  4880.3$\pm$2.7  &  4182$\pm$327  &  9.91$\pm$0.79  &  
	4860.8$\pm$1.8  &  1856$\pm$139  &  8.21$\pm$0.73  &  
	5007.9$\pm$0.2  &  349$\pm$12  &  6.62$\pm$0.24  &  
	0.76$\pm$0.26  &  1.79$\pm$0.59  &  2.94$\pm$0.56  \\
   &  6580.8$\pm$2.4  &  3349$\pm$97  &  33.65$\pm$2.03  &  
	6554.4$\pm$2.5  &  1487$\pm$139  &  27.91$\pm$0.98  &     &     &     &     &     &     \\
49963  &  4862.9$\pm$2.4  &  4208$\pm$416  &  7.92$\pm$2.11  &  
	4864.8$\pm$1.5  &  2320$\pm$176  &  10.92$\pm$2.10  &  
	5007.1$\pm$0.2  &  341$\pm$11  &  6.79$\pm$0.23  &  
	0.89$\pm$0.24  &  2.84$\pm$0.58  &     \\
   &  6559.3$\pm$2.4  &  2961$\pm$271  &  24.03$\pm$6.37  &  
	6561.9$\pm$2.1  &  1633$\pm$176  &  33.14$\pm$2.84  &     &     &     &     &     &     \\
49983  &  4892.1$\pm$3.2  &  4272$\pm$280  &  8.38$\pm$0.61  &  
	4858.8$\pm$1.5  &  1871$\pm$113  &  10.24$\pm$0.57  &  
	5007.3$\pm$0.2  &  325$\pm$11  &  6.47$\pm$0.24  &  
	0.56$\pm$0.25  &  3.72$\pm$0.51  &     \\
   &  6594.7$\pm$3.8  &  3468$\pm$91  &  26.81$\pm$1.65  &  
	6549.9$\pm$2.2  &  1519$\pm$113  &  32.76$\pm$0.77  &     &     &     &     &     &     \\
49990  &  4875.9$\pm$2.6  &  4628$\pm$301  &  7.65$\pm$0.77  &  
	4862.7$\pm$1.4  &  2146$\pm$113  &  11.48$\pm$0.78  &  
	5009.0$\pm$0.2  &  370$\pm$13  &  6.47$\pm$0.23  &  
	0.71$\pm$0.26  &  2.39$\pm$0.71  &  0.01$\pm$0.59  \\
   &  6580.0$\pm$2.8  &  3352$\pm$145  &  26.42$\pm$2.56  &  
	6562.3$\pm$1.9  &  1555$\pm$113  &  39.66$\pm$1.05  &     &     &     &     &     &     \\
50049  &  4874.9$\pm$2.3  &  4666$\pm$298  &  9.72$\pm$1.06  &  
	4864.2$\pm$1.6  &  2240$\pm$132  &  10.19$\pm$1.04  &  
	5007.2$\pm$0.2  &  348$\pm$11  &  6.59$\pm$0.23  &  
	0.82$\pm$0.23  &  2.08$\pm$0.72  &  0.08$\pm$0.57  \\
   &  6572.2$\pm$2.3  &  3001$\pm$136  &  29.72$\pm$3.05  &  
	6557.8$\pm$2.1  &  1441$\pm$132  &  31.16$\pm$1.40  &     &     &     &     &     &     \\
50052  &  4884.3$\pm$4.1  &  4710$\pm$315  &  6.07$\pm$0.78  &  
	4863.5$\pm$1.2  &  2214$\pm$96  &  15.64$\pm$0.83  &  
	5008.3$\pm$0.2  &  407$\pm$13  &  7.84$\pm$0.26  &  
	0.75$\pm$0.31  &  2.29$\pm$0.75  &  2.44$\pm$0.73  \\
   &  6590.2$\pm$5.2  &  3618$\pm$196  &  19.59$\pm$2.51  &  
	6562.1$\pm$1.7  &  1701$\pm$96  &  50.55$\pm$1.13  &     &     &     &     &     &     \\
50079  &  4879.9$\pm$2.8  &  4642$\pm$296  &  6.93$\pm$0.74  &  
	4865.8$\pm$1.3  &  2118$\pm$98  &  12.33$\pm$0.77  &  
	5009.3$\pm$0.1  &  343$\pm$11  &  6.94$\pm$0.22  &  
	0.68$\pm$0.24  &  1.83$\pm$0.63  &  0.05$\pm$0.51  \\
   &  6580.2$\pm$3.3  &  3606$\pm$180  &  24.17$\pm$2.50  &  
	6561.2$\pm$1.7  &  1645$\pm$98  &  43.01$\pm$1.04  &     &     &     &     &     &     \\
50094  &  4881.0$\pm$2.5  &  4778$\pm$286  &  7.66$\pm$0.68  &  
	4865.7$\pm$1.3  &  2165$\pm$104  &  11.87$\pm$0.69  &  
	5009.8$\pm$0.2  &  405$\pm$13  &  6.76$\pm$0.23  &  
	0.79$\pm$0.26  &  1.54$\pm$0.69  &  0.44$\pm$0.60  \\
   &  6582.3$\pm$2.7  &  3453$\pm$133  &  26.09$\pm$2.15  &  
	6561.7$\pm$1.8  &  1565$\pm$104  &  40.47$\pm$0.93  &     &     &     &     &     &     \\
50097  &  4880.6$\pm$2.7  &  5178$\pm$292  &  6.67$\pm$0.58  &  
	4863.5$\pm$1.2  &  2226$\pm$96  &  13.32$\pm$0.62  &  
	5008.3$\pm$0.2  &  387$\pm$13  &  6.75$\pm$0.22  &  
	0.79$\pm$0.26  &  0.81$\pm$0.61  &     \\
   &  6584.9$\pm$3.2  &  3633$\pm$153  &  22.28$\pm$1.83  &  
	6561.8$\pm$1.7  &  1562$\pm$96  &  44.52$\pm$0.84  &     &     &     &     &     &     \\
50110  &  4874.8$\pm$2.4  &  4758$\pm$304  &  9.18$\pm$0.67  &  
	4860.7$\pm$1.5  &  1949$\pm$113  &  9.94$\pm$0.66  &  
	5009.4$\pm$0.2  &  398$\pm$13  &  7.08$\pm$0.24  &  
	0.95$\pm$0.29  &  1.46$\pm$0.73  &  3.21$\pm$0.67  \\
   &  6578.2$\pm$2.4  &  3680$\pm$129  &  31.26$\pm$1.94  &  
	6559.2$\pm$2.2  &  1508$\pm$113  &  33.83$\pm$0.89  &     &     &     &     &     &     \\
\hline
\end{tabular}
\end{table}
\end{landscape}

\clearpage

\begin{landscape}
\begin{table}
\addtocounter{table}{-1}
\caption{continued}
\begin{tabular}{ccccccccccccc}
\hline
\multicolumn{1}{c}{JD} &\multicolumn{3}{c}{Inner Broad} & \multicolumn{3}{c}{Intermediate Broad}& \multicolumn{3}{c}{[O~{\sc iii}]} & \multicolumn{3}{c}{' '}\\
  & $W_0$ & $\sigma$ & flux & $W_0$ & $\sigma$ & flux & $W_0$ & $\sigma$ &
   flux & flux(H$\beta_N$) & flux(H$\alpha_N$) & flux([N~{\sc ii}]) \\
\hline
50333  &  4880.6$\pm$3.4  &  4200$\pm$282  &  6.32$\pm$0.73  &  
	4861.7$\pm$1.3  &  1984$\pm$99  &  12.32$\pm$0.77  &  
	5008.1$\pm$0.2  &  392$\pm$15  &  6.14$\pm$0.23  &  
	0.57$\pm$0.29  &  2.69$\pm$0.72  &  1.92$\pm$0.64  \\
   &  6585.6$\pm$4.1  &  3457$\pm$166  &  22.51$\pm$2.52  &  
	6560.1$\pm$1.7  &  1633$\pm$99  &  43.89$\pm$1.04  &     &     &     &     &     &     \\
50371  &  4864.0$\pm$2.1  &  3451$\pm$382  &  8.31$\pm$2.59  &  
	4862.8$\pm$1.2  &  1923$\pm$148  &  12.89$\pm$2.59  &  
	5007.1$\pm$0.2  &  340$\pm$11  &  7.59$\pm$0.26  &  
	0.69$\pm$0.31  &  2.06$\pm$0.68  &     \\
   &  6564.1$\pm$2.4  &  2710$\pm$285  &  24.94$\pm$7.76  &  
	6562.5$\pm$1.6  &  1510$\pm$148  &  38.73$\pm$3.50  &     &     &     &     &     &     \\
50391  &  4878.3$\pm$3.1  &  4352$\pm$307  &  6.66$\pm$0.81  &  
	4862.8$\pm$1.3  &  2030$\pm$107  &  13.21$\pm$0.83  &  
	5009.3$\pm$0.2  &  396$\pm$15.1  &  6.19$\pm$0.23  &  
	0.68$\pm$0.31  &  3.65$\pm$0.71  &     \\
   &  6583.1$\pm$3.8  &  3430$\pm$178  &  21.50$\pm$2.52  &  
	6562.2$\pm$1.7  &  1600$\pm$107  &  42.65$\pm$1.13  &     &     &     &     &     &     \\
50400  &  4863.3$\pm$1.6  &  3072$\pm$202  &  16.67$\pm$2.01  &  
	4867.3$\pm$1.6  &  1633$\pm$195  &  6.68$\pm$1.97  &  
	5008.9$\pm$0.2  &  381$\pm$13  &  7.93$\pm$0.27  &  
	0.75$\pm$0.34  &     &     \\
   &  6562.2$\pm$1.2  &  2570$\pm$109  &  51.71$\pm$6.01  &  
	6567.6$\pm$2.2  &  1366$\pm$195  &  20.72$\pm$2.66  &     &     &     &     &     &     \\
50401  &  4868.5$\pm$1.7  &  3959$\pm$255  &  12.45$\pm$0.96  &  
	4864.0$\pm$1.3  &  1696$\pm$113  &  9.99$\pm$0.93  &  
	5008.7$\pm$0.2  &  317$\pm$12  &  6.75$\pm$0.25  &  
	0.68$\pm$0.32  &  1.49$\pm$0.66  &     \\
   &  6564.7$\pm$1.5  &  2892$\pm$112  &  36.73$\pm$2.64  &  
	6558.6$\pm$1.7  &  1238$\pm$113  &  29.47$\pm$1.25  &     &     &     &     &     &     \\
50459  &  4880.7$\pm$1.9  &  4019$\pm$205  &  15.68$\pm$0.77  &  
	4857.6$\pm$1.3  &  1391$\pm$101  &  8.06$\pm$0.63  &  
	5010.9$\pm$0.1  &  322$\pm$10  &  6.42$\pm$0.25  &  
	0.57$\pm$0.28  &  6.89$\pm$0.72  &  7.31$\pm$0.49  \\
   &  6570.8$\pm$1.8  &  2722$\pm$68  &  32.96$\pm$1.29  &  
	6539.7$\pm$1.8  &  942$\pm$101  &  16.95$\pm$0.85  &     &     &     &     &     &     \\
50716  &  4868.0$\pm$1.4  &  3258$\pm$182  &  13.59$\pm$1.03  &  
	4863.5$\pm$1.1  &  1517$\pm$115  &  7.71$\pm$0.97  &  
	5009.3$\pm$0.1  &  320$\pm$11 &  6.25$\pm$0.22  &  
	0.62$\pm$0.28  &  3.43$\pm$0.73  &     \\
   &  6571.2$\pm$1.1  &  2660$\pm$86  &  45.12$\pm$3.23  &  
	6565.1$\pm$1.6  &  1238$\pm$115  &  25.58$\pm$1.31  &     &     &     &     &     &     \\
50717  &  4863.0$\pm$2.1  &  3559$\pm$283  &  7.51$\pm$1.46  &  
	4867.2$\pm$1.3  &  1928$\pm$126  &  9.74$\pm$1.47  &  
	5004.2$\pm$0.1  &  322$\pm$9  &  7.79$\pm$0.24  &  
	1.06$\pm$0.24  &     &  0.08$\pm$0.51  \\
   &  6557.3$\pm$2.8  &  2936$\pm$283  &  29.94$\pm$1.98  &  
	6563.0$\pm$1.2  &  1591$\pm$91  &  38.82$\pm$5.82  &     &     &     &     &     &     \\
50747  &  4867.9$\pm$1.4  &  4157$\pm$224  &  13.93$\pm$0.91  &  
	4867.6$\pm$1.1  &  1778$\pm$106  &  9.77$\pm$0.88  &  
	5008.5$\pm$0.1  &  347$\pm$11  &  7.14$\pm$0.23  &  
	0.94$\pm$0.27  &     &  0.88$\pm$0.60  \\
   &  6567.3$\pm$1.1  &  2819$\pm$88  &  40.16$\pm$2.34  &  
	6566.9$\pm$1.6  &  1205$\pm$106  &  28.16$\pm$1.19  &     &     &     &     &     &     \\
50771  &  4861.4$\pm$1.2  &  3315$\pm$180  &  16.79$\pm$0.61  &  
	4857.3$\pm$1.4  &  1017$\pm$101  &  3.69$\pm$0.63  &  
	5006.0$\pm$0.2  &  439$\pm$13  &  7.56$\pm$0.24  &  
	0.27$\pm$0.40  &  2.01$\pm$1.73  &     \\
   &  6560.4$\pm$0.6  &  2182$\pm$39  &  48.98$\pm$1.23  &  
	6554.9$\pm$1.9  &  669$\pm$101  &  10.78$\pm$0.85  &     &     &     &     &     &     \\
50773  &  4863.1$\pm$1.2  &  3232$\pm$139  &  17.14$\pm$0.64  &  
	4870.2$\pm$1.2  &  861$\pm$123  &  3.27$\pm$0.45  &  
	5011.7$\pm$0.1  &  298$\pm$8  &  8.07$\pm$0.25  &  
	0.78$\pm$0.28  &     &  5.46$\pm$0.81  \\
   &  6560.1$\pm$0.7  &  2242$\pm$36  &  48.95$\pm$1.03  &  
	6569.6$\pm$1.6  &  597$\pm$123  &  9.35$\pm$0.60  &     &     &     &     &     &     \\
50806  &  4866.6$\pm$1.5  &  3543$\pm$214  &  11.04$\pm$1.12  &  
	4868.4$\pm$1.2  &  1720$\pm$115  &  8.82$\pm$1.09  &  
	5010.2$\pm$0.1  &  340$\pm$11  &  6.89$\pm$0.23  &  
	0.75$\pm$0.26  &     &     \\
   &  6565.1$\pm$1.2  &  2782$\pm$115  &  39.52$\pm$3.77  &  
	6567.5$\pm$1.6  &  1350$\pm$115  &  31.55$\pm$1.47  &     &     &     &     &     &     \\
50831  &  4860.8$\pm$2.9  &  4669$\pm$419  &  5.902$\pm$1.20  &  
	4868.0$\pm$1.3  &  2380$\pm$115  &  14.53$\pm$1.23  &  
	5011.1$\pm$0.2  &  381$\pm$11  &  7.41$\pm$0.23  &  
	1.41$\pm$0.27  &     &     \\
   &  6560.3$\pm$3.4  &  3178$\pm$260  &  19.91$\pm$3.98  &  
	6569.9$\pm$1.8  &  1620$\pm$115  &  49.00$\pm$1.67  &     &     &     &     &     &     \\
51051  &  4871.9$\pm$1.5  &  3641$\pm$214  &  13.83$\pm$0.83  &  
	4864.7$\pm$1.2  &  1418$\pm$108  &  7.45$\pm$0.79  &  
	5007.8$\pm$0.2  &  368$\pm$12  &  6.81$\pm$0.22  &  
	0.43$\pm$0.30  &  2.49$\pm$1.01  &  0.56$\pm$0.82  \\
   &  6564.5$\pm$1.0  &  2708$\pm$65  &  41.72$\pm$1.72  &  
	6554.7$\pm$1.7  &  1055$\pm$108  &  22.48$\pm$1.07  &     &     &     &     &     &     \\
51074  &  4862.5$\pm$2.4  &  3658$\pm$421  &  5.43$\pm$1.91  &  
	4864.7$\pm$1.1  &  2092$\pm$117  &  13.19$\pm$1.92  &  
	5004.1$\pm$0.2  &  414$\pm$12  &  7.56$\pm$0.22  &  
	0.82$\pm$0.27  &     &     \\
   &  6557.7$\pm$3.3  &  2946$\pm$421  &  19.73$\pm$2.57  &  
	6560.6$\pm$0.8  &  1685$\pm$81  &  48.01$\pm$7.02  &     &     &     &     &     &     \\
51084  &  4861.4$\pm$1.3  &  2797$\pm$108  &  13.12$\pm$0.42  &  
	4865.2$\pm$0.9  &  797$\pm$51  &  3.52$\pm$0.27  &  
	5008.1$\pm$0.1  &  345$\pm$10  &  7.35$\pm$0.22  &     &     &     \\
   &  6561.4$\pm$1.3  &  2259$\pm$108  &  57.19$\pm$0.56  &  
	6566.5$\pm$0.7  &  644$\pm$35  &  15.34$\pm$1.12  &     &     &     &     &     &     \\
\hline
\end{tabular}
\end{table}
\end{landscape}

\clearpage

\begin{figure*}
\centering\includegraphics[width = 18cm,height=22cm]{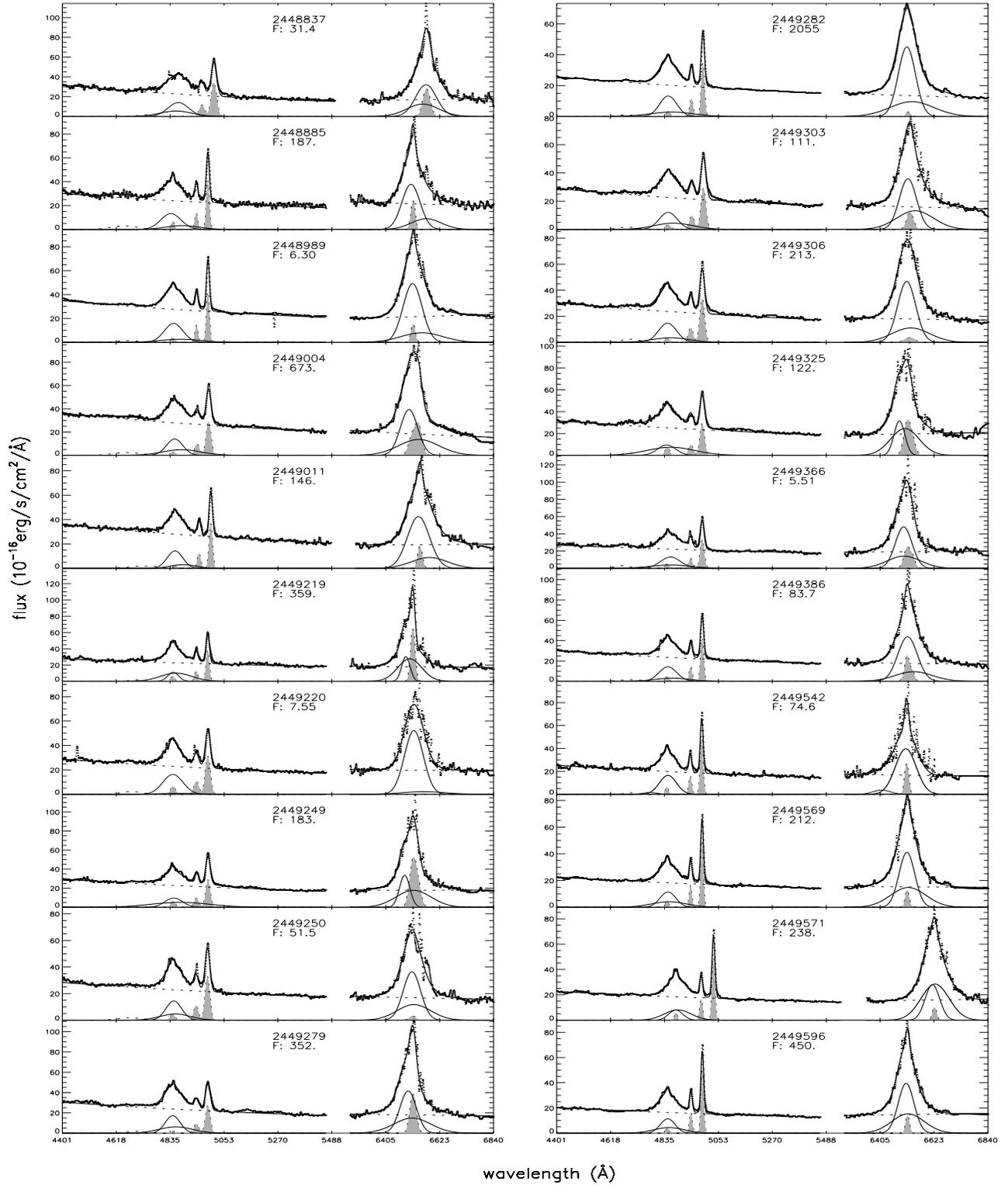}
\caption{Best fitted results for the lines around the H$\beta$ and around 
the H$\alpha$ under the model with two broad components applied for each 
observed broad balmer line. The dotted lines are for the observed spectra, 
the thick solid lines are for the best fitted results, the AGN continuum 
is shown in the dotted line under the spectrum. The thick and the thin 
solid lines near the bottom represent the two broad components for each 
observed broad balmer line, the shadow areas near the bottom show the 
narrow lines, the dotted line near the bottom shows the He~{\sc ii} line. 
In each panel, the information of the MJD (seven characters) and the F 
value (starting with F:) calculated by the equation (6) are marked. 
}
\label{hab2_fit}
\end{figure*}
\setcounter{figure}{3}
\begin{figure*}
\centering\includegraphics[width = 18cm,height=22cm]{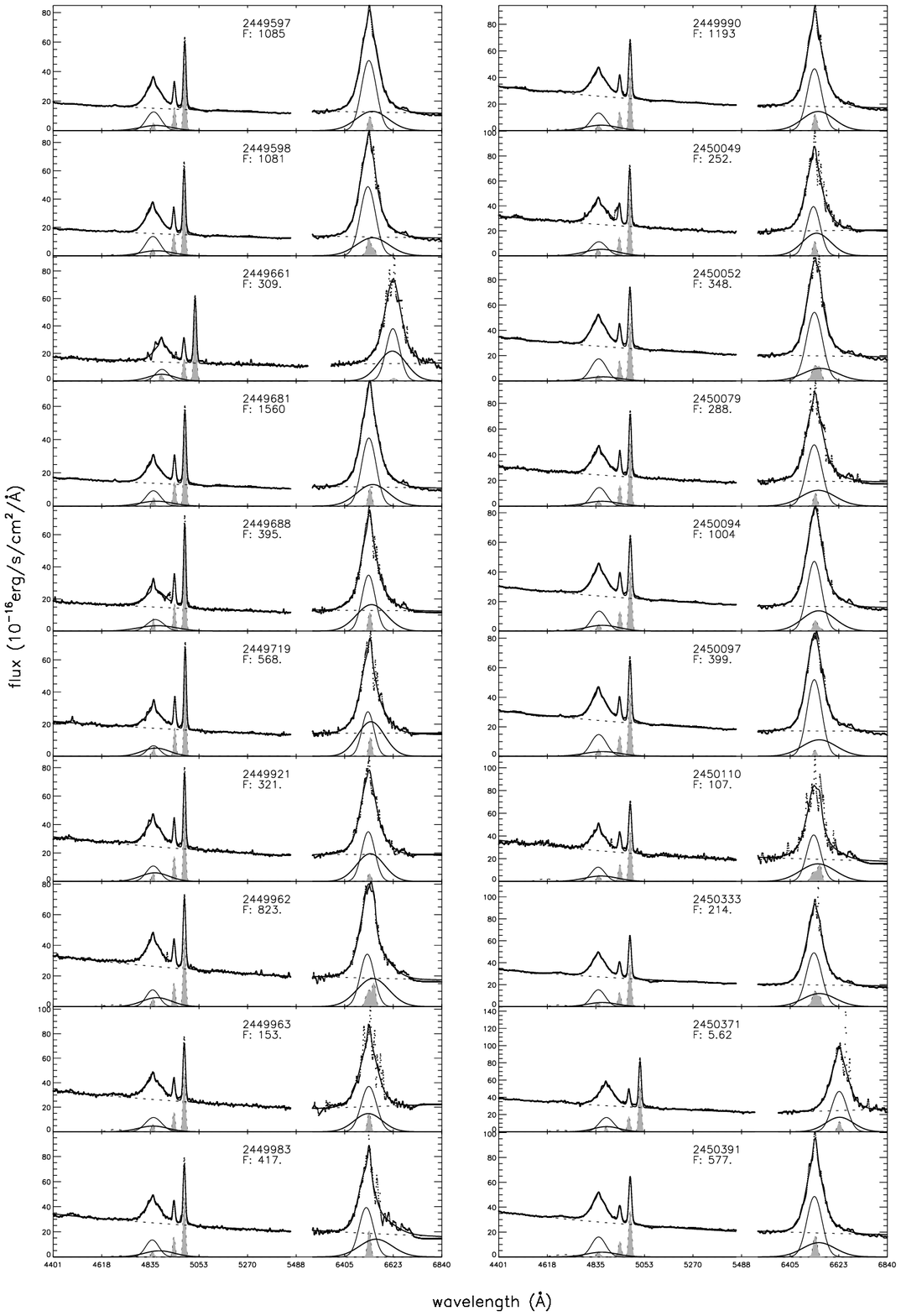}
\caption{--continued.
}
\end{figure*}
\setcounter{figure}{3}
\begin{figure*}
\centering\includegraphics[width = 18cm,height=20cm]{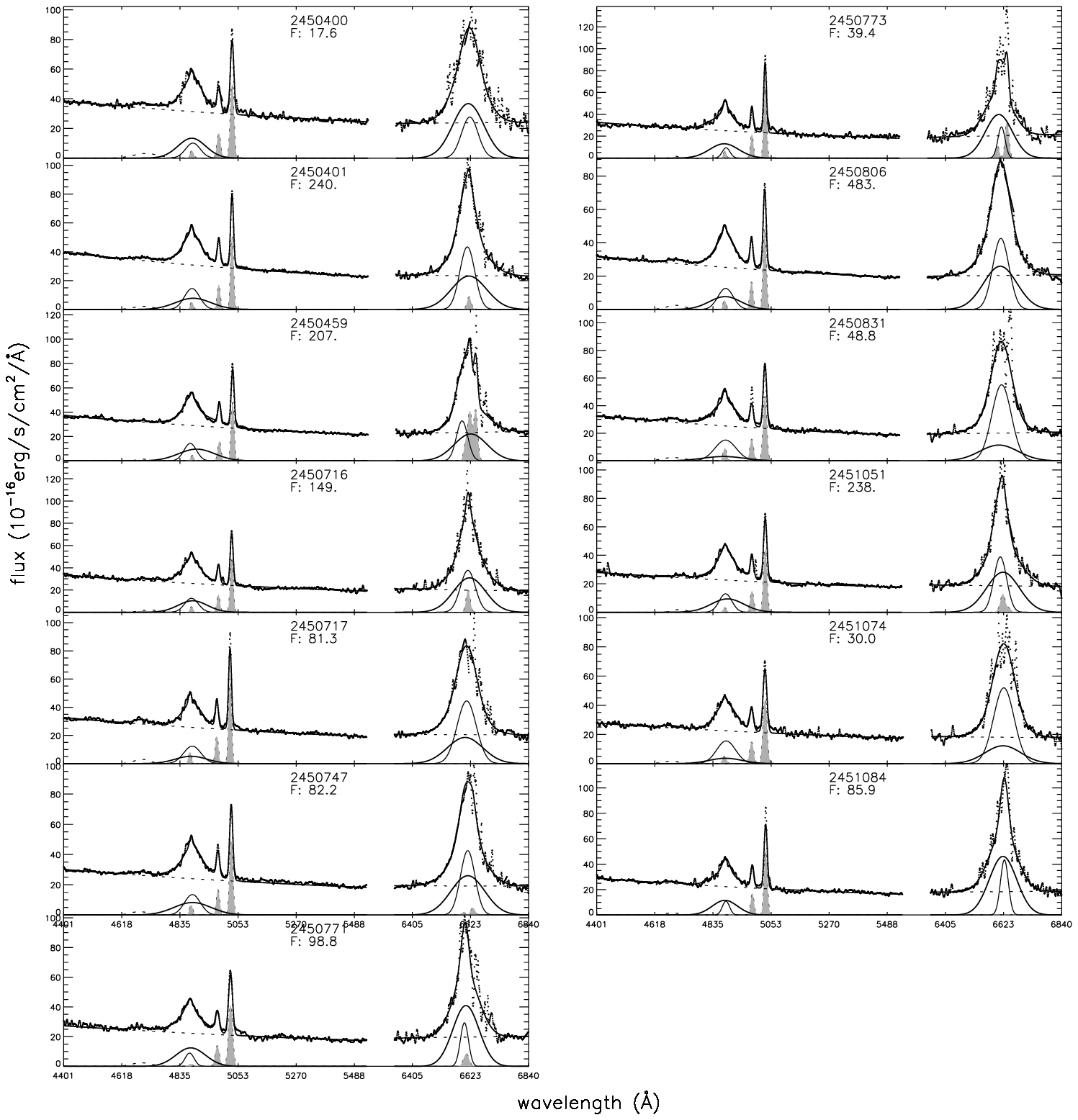}
\caption{--continued.
}
\end{figure*}

\begin{figure*}
\centering\includegraphics[width = 18cm,height=10cm]{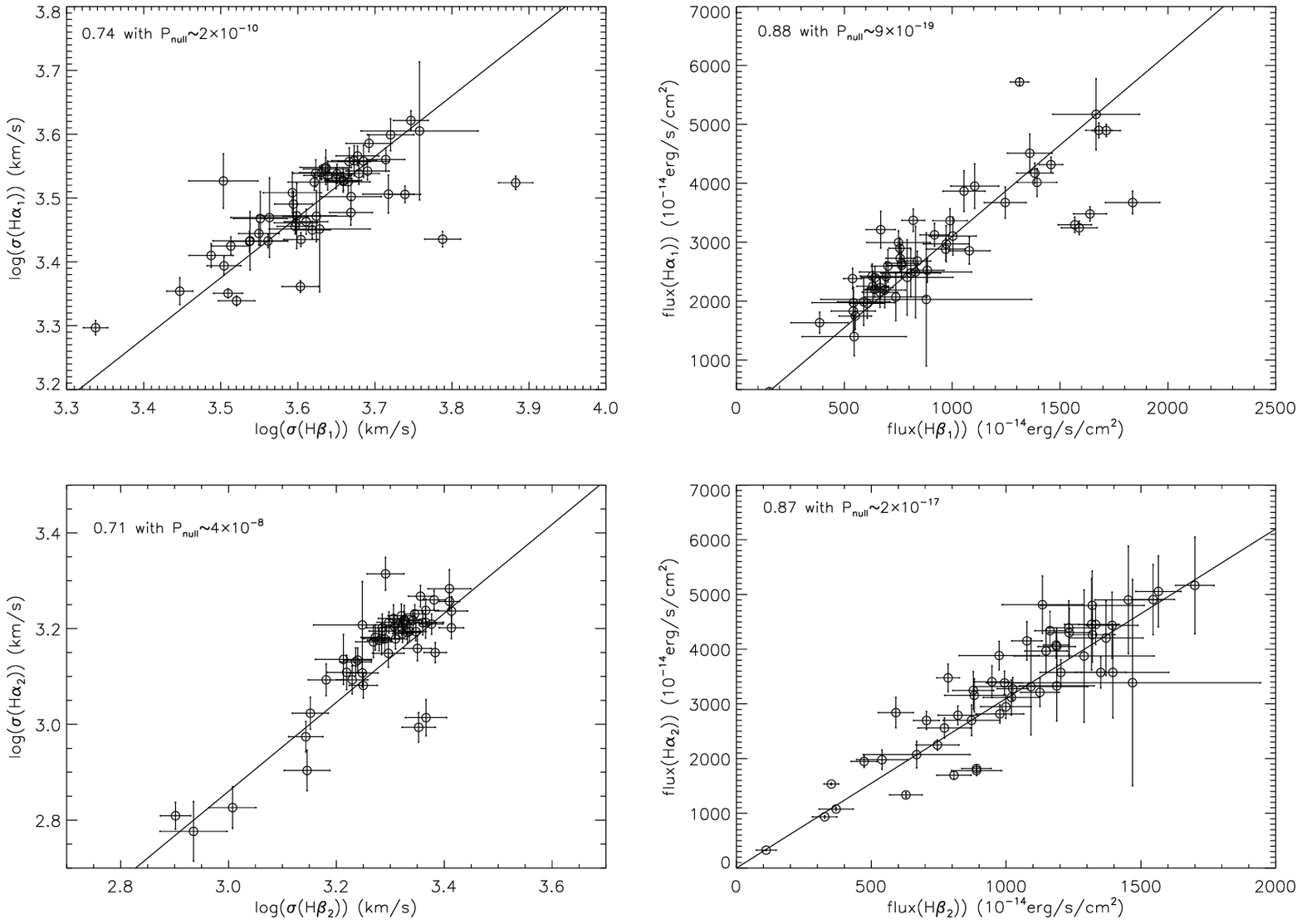}
\caption{The broad line width correlation (the left panels) and 
the broad line flux correlation (the right panels) between 
the two broad components of the H$\alpha$ and the H$\beta$. The 
solid lines are the best fitted results for the correlations. The 
coefficient and the corresponding $P_{null}$ are marked in each panel 
for the correlation.}
\label{hab2}
\end{figure*}

\begin{figure*}
\centering\includegraphics[width = 10cm,height=8cm]{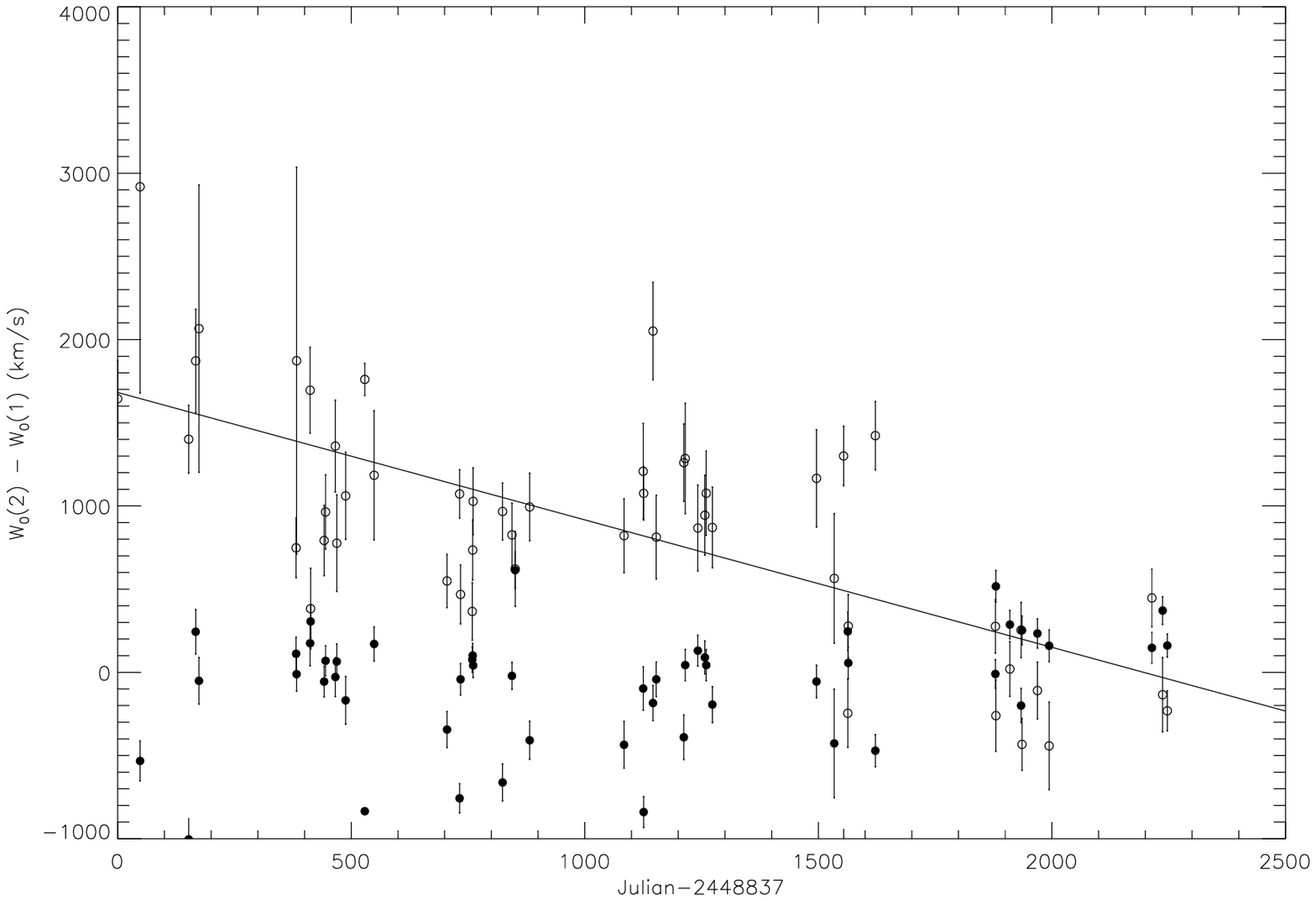}
\caption{ The properties of the relative shifted velocities of the inner 
broad component to the intermediate broad component (the open 
circles), and the relative shifted velocities of the intermediate 
broad component to the narrow [O~{\sc iii}]$\lambda5007$\AA (the 
solid circles). The solid line shows the time dependence of the 
relative shifted velocities of the inner broad component to the 
intermediate broad component.
}
\label{vel}
\end{figure*}

\iffalse
\begin{figure*}
\centering\includegraphics[width = 10cm,height=8cm]{Figs/check_o3.ps}
\caption{ The properties of the [O~{\sc iii}]$\lambda5007$\AA flux. 
}
\label{o3}
\end{figure*}
\fi

\begin{figure*}
\centering\includegraphics[width = 18cm,height=9cm]{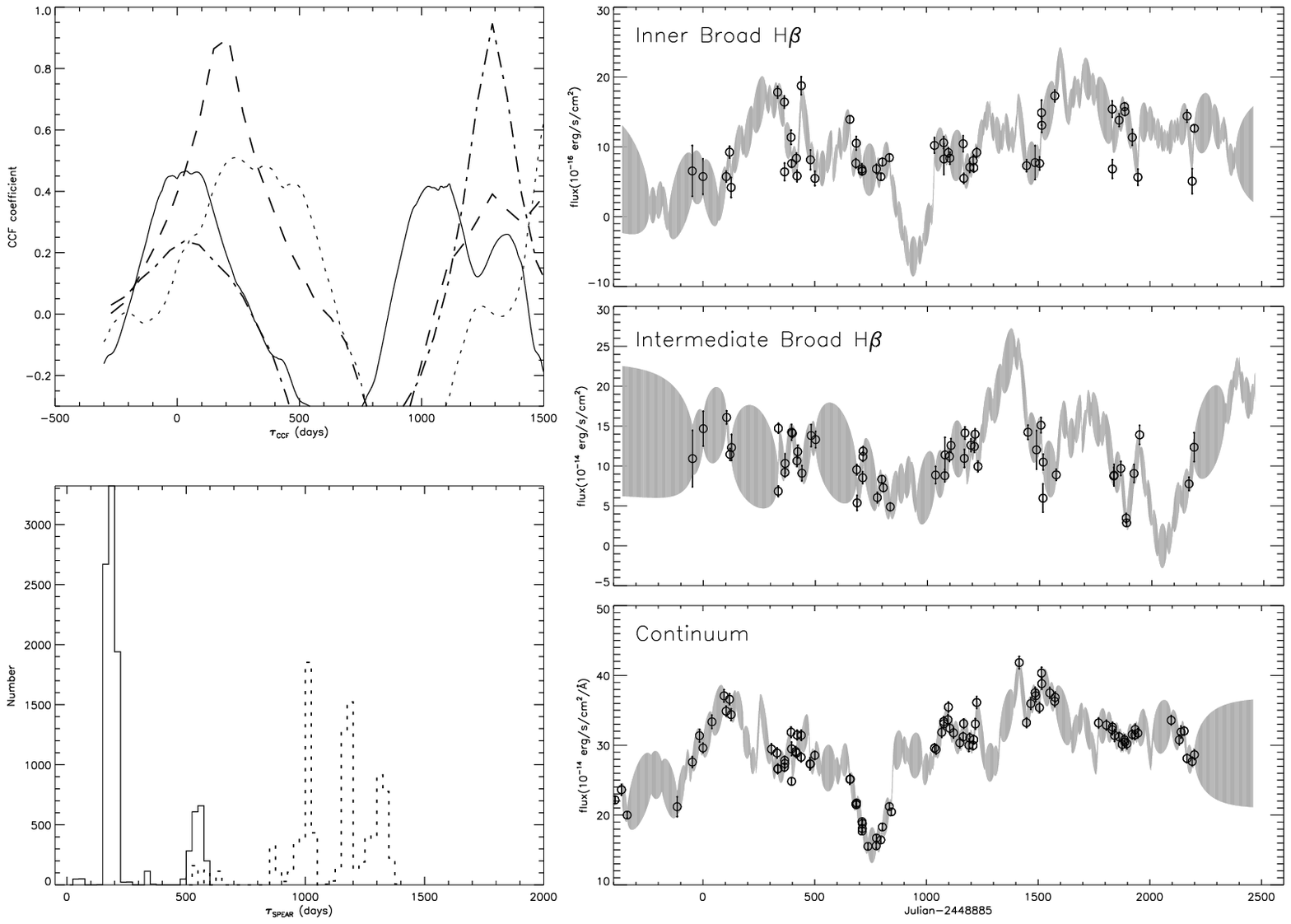}
\caption{The ICCF and SPEAR results for the inner and the intermediate
broad components of the H$\beta$ of PG 0052+251. Top left panel 
shows the ICCF results for the inner broad H$\beta$ (dotted 
line for the result through the observational light curves,
dashed line for the result through the light curves determined by the
random walk model) and the intermediate broad H$\beta$ (solid line
for the result through the observational light curves, dot-dashed line for
the result through the light curves determined by the random walk model),
and bottom left panel shows the distributions of the time lags determined by
the SPEAR method (thin solid line for H$\beta_1$ and thick dotted
line for H$\beta_2$). The right panels from top 
to bottom shows the observational light curves (open circles) of 
the inner broad H$\beta$, the intermediate broad H$\beta$ and the continuum 
emission, and the corresponding best descriptions for the observational 
light curves by the damped random walk model for AGN variability. 
}
\label{lag_hb}
\end{figure*}

\begin{figure*}
\centering\includegraphics[width = 18cm,height=9cm]{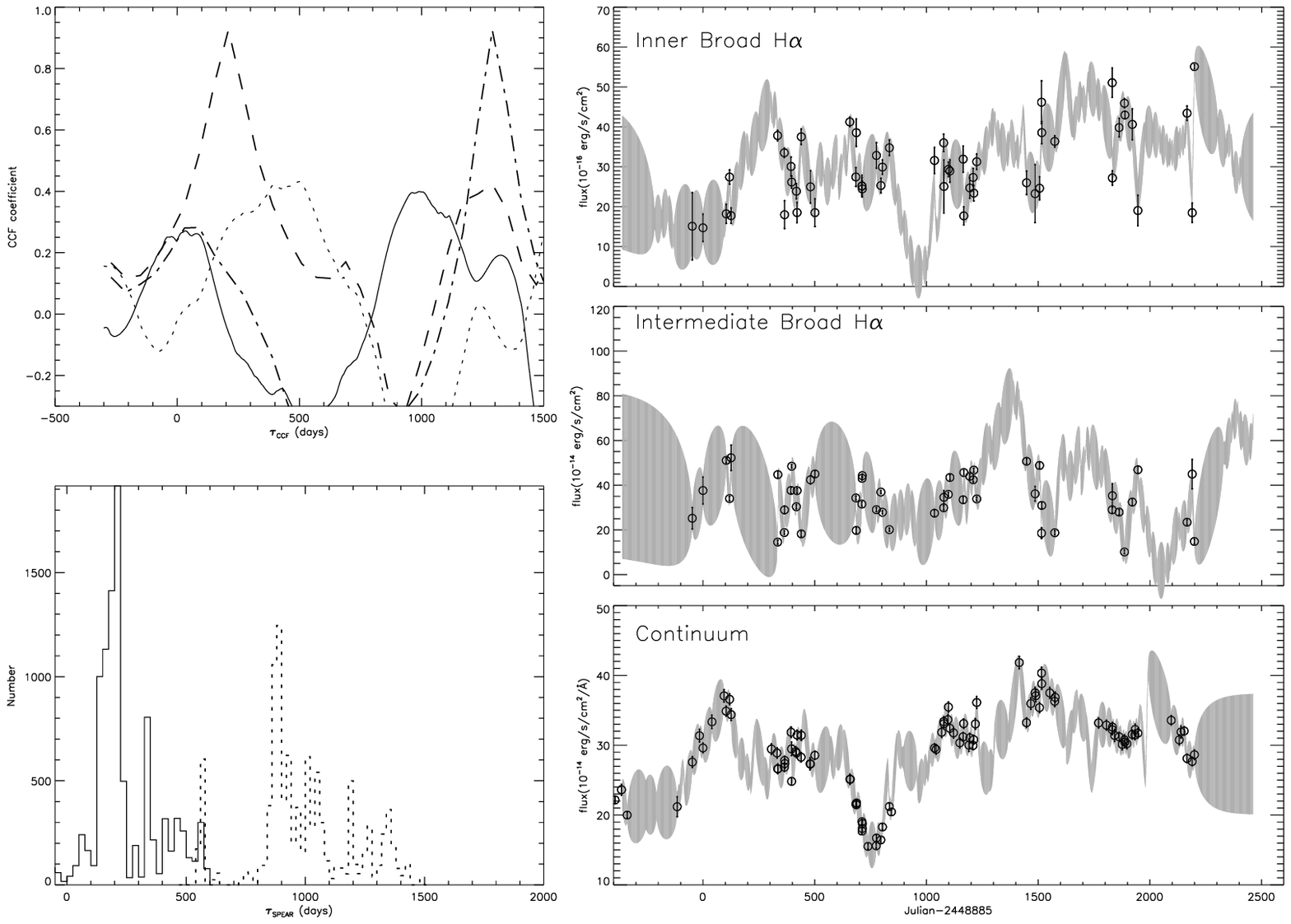}
\caption{The ICCF and SPEAR results for the H$\alpha$, symbols and lines 
have the same meanings as those in the Figure~\ref{lag_hb}, but for 
the H$\alpha$.
}
\label{lag_ha}
\end{figure*}

\begin{figure*}
\centering\includegraphics[width = 18cm,height=8cm]{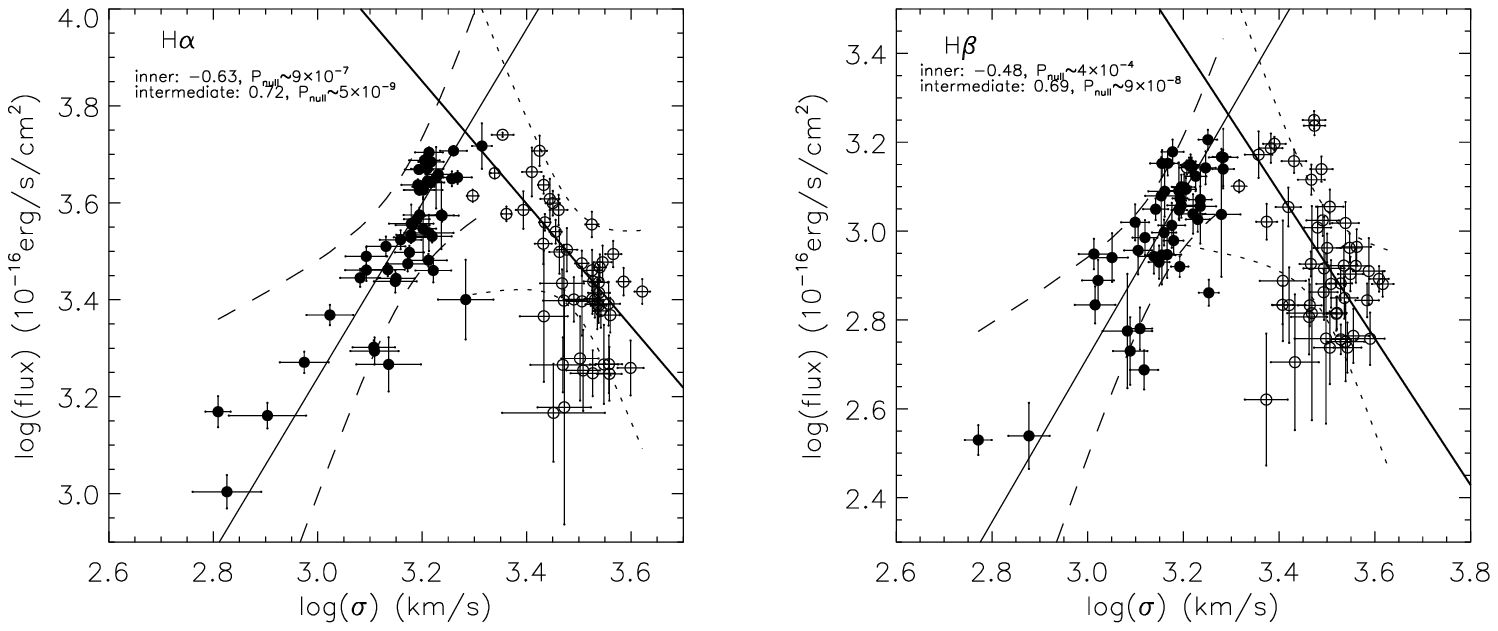}
\caption{On the correlations between the broad line flux and the broad line 
width for the inner broad component, the intermediate broad component of 
the H$\alpha$ (the left panel) and the H$\beta$ (the right 
panel). The open circles are for the inner broad components, the solid 
circles are for the intermediate broad components. The thick solid line, 
the thin solid line, the dotted lines and the dashed lines show the best 
fitted results for the correlations for the inner broad components, for 
the intermediate broad components, and the corresponding  99.95\% 
confidence bands for the best fitted results.}
\label{flux_width}
\end{figure*}

\end{document}